\documentclass[aps,prd,twocolumn,superscriptaddress,nofootinbib,10pt,reprint,preprintnumbers]{revtex4-1}

\usepackage{graphicx,amsmath,amsfonts,amssymb,aas_macros,slashed}
\usepackage{bbold}
\usepackage{datetime}
\usepackage{numprint}
\usepackage{enumitem}
\usepackage{booktabs}

\usepackage{graphicx,color,amsmath}
\usepackage{hyperref}

\allowdisplaybreaks

\newcolumntype{C}{>{$}c<{$}}
\AtBeginDocument{
\heavyrulewidth=.08em
\lightrulewidth=.05em
\cmidrulewidth=.03em
\belowrulesep=.65ex
\belowbottomsep=0pt
\aboverulesep=.4ex
\abovetopsep=0pt
\cmidrulesep=\doublerulesep
\cmidrulekern=.5em
\defaultaddspace=.5em
}

\newcommand{\eV}{\ensuremath{\mathrm{eV}}}

\newcommand{\MeV}{\ensuremath{\mathrm{MeV}}}
\newcommand{\GeV}{\ensuremath{\mathrm{GeV}}}

\usepackage{pgfplots}
\usepackage{bm}
\usepackage{mathtools}

\begin{document}

\preprint{UCI-HEP-TR-2022-12}

\title{Shining Light on Cosmogenic Axions with Neutrino Experiments}

\author{Yanou Cui}
\email{yanou.cui@ucr.edu}
\affiliation{Department of Physics and Astronomy, University of California, Riverside, CA 92521, USA}

\author{Jui-Lin Kuo}
\email{juilink1@uci.edu}
\affiliation{Department of Physics and Astronomy, University of California, Irvine, CA 92697-4575, USA}

\author{Josef Pradler}
\email{josef.pradler@oeaw.ac.at}
\affiliation{Institute of High Energy Physics, Austrian Academy of Sciences, Georg-Coch-Platz 2, 1010 Vienna, Austria}
\affiliation{CERN, Theoretical Physics Department, 1211 Geneva 23, Switzerland}

\author{Yu-Dai Tsai}
\email{yudait1@uci.edu}\email{yt444@cornell.edu}
\affiliation{Department of Physics and Astronomy, University of California, Irvine, CA 92697-4575, USA}

\begin{abstract}
While most searches for cosmic axions so far focused on their cold relics as (a component of) dark matter, various well-motivated cosmological sources can produce ``boosted'' axions that remain relativistic today. We demonstrate that existing/upcoming neutrino experiments such as Super-Kamiokande, Hyper-Kamiokande, DUNE, JUNO, and IceCube can probe such energetic axion relics. The characteristic signature is the mono-energetic single photon signal from axion absorption induced by the axion-photon coupling. This proposal offers to cover parameter ranges complementary to existing axion searches and provides new opportunities for discovery with neutrino facilities.
\end{abstract}

\maketitle

\section{Introduction}
\label{sec:introduction}
Axion-like particles (ALPs) are predicted in many well-motivated beyond the Standard Model (BSM) particle physics theories and can be compelling dark matter candidates that attracted increasing interest in recent years. The relics of stable ALPs naturally form a cold, cosmological background~\cite{Preskill:1982cy,Abbott:1982af,Dine:1982ah}, and have been the primary target for ALP searches. On the other hand, there are many well-motivated sources producing a cosmic background of energetic axions, \textit{e.g.}, dark matter decay, supernovae, topological defects' radiation and parametric resonance. In this \textit{Letter} we explore the prospects of detecting cosmologically produced ALPs with large volume neutrino detectors, which are well-suited for ALPs of $\mathcal{O}({\rm MeV})$ or higher energy. The smoking-gun signal would be a single mono-energetic photon event resulting from the Inverse-Primackoff (IP) process.

A number of recent literature consider related topics, yet cover different bases. For instance, a recent study on searches for energetic cosmological axions can be found in~\cite{Dror:2021nyr}, but the energy scales of axions they focused on are lower than our interested regime, with different relevant experiments. Axion detection via absorption has also been studied in a context of dark matter (DM) direct detection experiments, which concerns cold relics with $\mathcal{O}({\rm keV})$ energies~\cite{Pospelov:2008jk, An:2014twa, Bloch:2016sjj}. Another recent study considered axion searches with the Deep Underground Neutrino Experiment (DUNE) based on IP process~\cite{Brdar:2020dpr}. However, they consider beam-produced axions which may decay within the detector, and thus cannot be DM candidates. In addition, the searches in~\cite{Brdar:2020dpr} are relevant for DUNE near detector, while the signal considered in this work is cosmogenic and thus is relevant for DUNE far detectors. Therefore, this \textit{Letter} well complements existing studies on axion searches with astrophysical or terrestrial probes~\cite{ADMX:2020ote,Garcon:2017ixh,Kahn:2018fgp,CAST:2017uph,IAXO:2019mpb,Spector:2016vwo}, and provides new, timely BSM search targets with potential DM connections for upcoming neutrino experiments such as Hyper-Kamiokande (HK)~\cite{Hyper-Kamiokande:2018ofw}, DUNE~\cite{DUNE:2020ypp} and Jiangmen Underground Neutrino Observatory (JUNO)~\cite{JUNO:2021vlw}. The spirit of the present study is similar to previous work by some of the authors, that explored the direct detection sensitivity of DM and neutrino experiments to a dark radiation (DR) flux in the form of neutrinos and sourced by DM decay~\cite{Cui:2017ytb,Nikolic:2020fom,Bondarenko:2020vta,Kuo:2021mtp,Gu:2021lni}.

The scenario considered here resonates with the idea of boosted DM (BDM) which has been pursued by many neutrino experimental collaborations~\cite{Agashe:2014yua, Berger:2014sqa, Giudice:2017zke, Berger:2019ttc, Huang:2013xfa, Kopp:2015bfa, Giudice:2017zke, Super-Kamiokande:2017dch, COSINE-100:2018ged, Chatterjee:2018mej, DUNE:2020ypp,Dent:2019krz, PandaX-II:2021kai}. However, the phenomenology and search strategies of the two are distinct from each other: BDM is typically thermal WIMP-like particles produced from processes such as annihilation or decay of DM, and the detection is based on the scattering process. In contrast, the DR-like ``boosted" ALPs considered here are very light, very weakly interacting, and would be detected by absorption processes. On another note, in principle, cold axion of mass $\mathcal{O}$(MeV) or heavier can be probed by neutrino detectors via IP process, in analogy to keV ALP searches with DM direct detection experiments. However, cosmogenic ALPs of such masses are generally too short-lived, which may be subject to strong constraints~\cite{Cadamuro:2012rm}, or have too small interaction rates for detection due to small fluxes and weak couplings.

In the following, we will first outline a benchmark example of energetic ALPs production from DM decay in Sec.~\ref{sec:flux}, and the interactions relevant to the experimental probes with neutrino detectors in Sec.~\ref{sec:detection}. Then in Sec.~\ref{sec:experiments} we will briefly review relevant information about a few representative neutrino experiments to be considered. Existing constraints and projection for sensitivities with future experiments will be shown following the analysis of the signal in Sec.~\ref{sec:results}. Finally, we will conclude in Sec.~\ref{sec:conclusions} .\\

\section{Flux of ALPs from dark matter decay}
\label{sec:flux}

Energetic ALPs can originate from various cosmological sources, such as DM decay, emission from topological defects and parametric resonance~\cite{Dror:2021nyr, Chang:2021afa}. As a benchmark example, in this work we consider the scenario where the relativistic ALPs are sourced from the two-body decay of DM $X\rightarrow aa$ with a unity branching fraction. For simplicity, we assume that the decaying DM constitutes $\sim100\%$ of DM. However, the cold relic of the ALP can be a non-negligible fraction of all DM, and the ALPs produced from $X$ decay can have an observable impact on cosmological structure formation, as suggested by earlier literature on boosted DM and dark radiation~\cite{Hasenkamp:2012ii,Cui:2018imi}. Here we focus on terrestrial probes of the ``boosted" ALPs with neutrino experiments, while other ramifications are worth future investigation.

We take a general approach insensitive to the specifics of the $Xaa$ interaction, and trade the $Xaa$ coupling for the lifetime of the progenitor, $\tau_X$. 
In the local Universe, the Galactic energy differential ALP flux can be expressed as~\cite{Cui:2017ytb}
\begin{align}
\label{eq:galflux}
\dfrac{d\Phi_a^{\rm gal}}{dE_a} = \dfrac{e^{-t_0/\tau_X}}{m_X \tau_X} \dfrac{dN_a}{dE_a} R_\odot \rho_\odot \langle D\rangle\,,
\end{align}
where $t_0$ is the age of the universe, $m_X$ is the DM mass, $R_\odot \simeq 8.33\,{\rm kpc}$ is the distance between the Sun and the Galactic center, $\rho_\odot \simeq 0.3\,{\rm GeV}/{\rm cm}^3$ is the local DM mass density and $\langle D\rangle$ is the sky-averaged $D$-factor; in the numerical evaluation, we take a NFW profile~\cite{Navarro:1995iw}. 
If only a fraction of DM is assumed to decay, all flux formulae are to be multiplied by such a fraction in a trivial way. 
With negligible correction by the velocity dispersion of $X$, the ALP spectrum per decay, $dN_a/dE_a$, can be approximated by the monochromatic form, which reads
\begin{align}
\dfrac{dN_a}{dE_a} = 2\,\delta\left(E_a - m_X/2 \right)\,,
\end{align}
where $\delta(E_a - m_X/2)$ is a Dirac delta function.

On the other hand, the extragalactic energy differential ALP flux is given by~\cite{Cui:2017ytb},
\begin{align}
\label{eq:egalflux}
\dfrac{d\Phi_a^{\rm ext}}{dE_a} = \dfrac{2\Omega_X \rho_c}{m_X \tau_X H_0 p_a} \dfrac{e^{-t(\xi -1)/\tau_X}}{\sqrt{\xi^3 \Omega_m +\Omega_\Lambda}} \Theta (\xi -1)\,,
\end{align}
with $p_a = \sqrt{E_a^2 -m_a^2}$ and $m_a$ being the ALP mass, $\Omega_X = 0.2607$ being the density parameter of DM~\cite{Planck:2018vyg}, $\rho_c = 4.82\times 10^{-6}\,{\rm GeV}\,{\rm cm}^{-3}$ being the critical density of the Universe today, $\Theta(\xi -1)$ being a Heaviside step function, and $\xi \equiv \sqrt{(m_X/2)^2 -m_a^2}/p_a$, respectively. 
Finally, $t(z)$ is the  cosmic time at redshift $z$ and we assume a standard cosmological expansion history with the matter density parameter $\Omega_m = 0.3111$, the dark energy density parameter is $\Omega_\Lambda = 0.6889$, and where we take the current Hubble expansion rate as $H_0 = 67.66\,{\rm km}\,{\rm s}^{-1}\,{\rm Mpc}^{-1}$~\cite{Planck:2018vyg}.

In this work, we take a benchmark DM lifetime of $\tau_X =35\,t_0$, which satisfies cosmological bounds on DM decaying solely into DR~\cite{DES:2020mpv} (see~\cite{Enqvist:2015ara,Poulin:2016nat,Nygaard:2020sow,Simon:2022ftd} for other studies about constraints on DM lifetime and fraction of decaying DM).
We note in passing that stimulated $X\to aa$ decay in the DR background given by~\eqref{eq:galflux} and~\eqref{eq:egalflux} remains negligible as the ALP occupation number stays well below unity for our parameters of interest.
As long as the ALP DR remains relativistic, the laboratory signatures considered are insensitive to the value of  $m_a$, thus we take it to zero in our calculations, unless stated otherwise. Therefore, effectively, the only free remaining parameter determining the axion flux is the mass of the progenitor~$m_X$. Astrophysical constraints, however, sensitively depend on $m_a$, and we will comment on it in the following section. This leaves us two free parameters in this framework to be studied: the mass of the progenitor $m_X$ and the coupling strength between ALP and SM photon $g_{\rm a \gamma \gamma}$.

Integrating over the ALP energy $E_a$ one obtains the total flux~$\Phi_a$. For example, taking  $m_X = 30\,{\rm MeV}$ yields $\Phi_a \sim 10^5\,{\rm cm}^{-2}{\rm s}^{-1}$, much smaller than  the total solar neutrino flux $\Phi_\nu^{\odot} \sim 10^{11}\,{\rm cm}^{-2}{\rm s}^{-1}$ (or the total ${}^8$B flux). Therefore, large volume neutrino detectors, as considered in this work, are favored for probing the wide parameter region with larger progenitor masses $m_X\gtrsim 30~$MeV. The resulting galactic and extragalactic energy-differential fluxes of ALP DR are illustrated in the Appendix (Fig.~\ref{fig:axion_flux}) and can also be found in \cite{Cui:2017ytb, Kuo:2021mtp}.

\section{ALP interaction and astrophysical constraints}
\label{sec:detection}
The ALP as considered can potentially be detected via its interaction with photons $\gamma$, $e^{-}$ or nucleons $N$. In addition to laboratory probes, all couplings are subject to strong constraints from astrophysics that typically span broad ranges in axion mass. The bound on the axion-electron coupling is particularly strong, such that it supersedes the detectability of the boosted ALPs with large volume neutrino experiments as considered in this work.
In contrast, as we shall see, the $a\gamma\gamma$ coupling is accessible in the laboratory with a comparatively clean signature (in comparison to the electron or nucleon signals considering the backgrounds).

In this work, we therefore focus on the coupling between the ALP and photon, $g_{a\gamma\gamma}$. The interaction is described by
\begin{align}
 \mathcal{L}= - \dfrac{1}{4} g_{a\gamma\gamma} a F^{\mu\nu} \tilde{F}_{\mu\nu}\,,
\end{align}
where $g_{a\gamma\gamma}$ has mass-dimension $-1$, $F^{\mu\nu}$ is the photon field strength and $\tilde{F}_{\mu\nu} = \epsilon_{\mu\nu\alpha\beta} F^{\alpha\beta}/2$.
The $g_{a\gamma\gamma}$ coupling induces the axion decay into a photon-pair with a decay rate at rest given by
\begin{align}
\label{eq:axionLifetime}
 \Gamma^{(0)}_{a\rightarrow \gamma\gamma} = \dfrac{g_{a\gamma\gamma}^2 m_a^3 }{64\pi} = \tau_a^{-1}\,,
\end{align}
with $m_a$ and $\tau_a$ being the mass and lifetime of the ALP.

The two-body decay of the boosted ALP DR into pairs of photons is severely constrained by indirect detection searches of decaying DM, see, \textit{e.g.}~\cite{Cadamuro:2011fd,Cohen:2016uyg,Foster:2021ngm,Roach:2022lgo}. Due to the large boost of the axions, the induced photon spectrum is essentially identical to the DR one, thus the galactic flux is constrained from line searches, and the cosmological flux from a continuum excess.
Let us denote by $\tau_{\rm min}(m_{\rm DM})$ the observationally inferred lifetime constraint from DM decay into a pair of  photons. Considering that each $X$-decay eventually yields four photons and $\tau_{\rm min}\gg \tau_X$, furthermore, neglecting redshift, we find the condition on the in-flight axion lifetime: $\gamma \tau_a \gtrsim 2 \tau_{\rm min}(m_X/2) $, where $\gamma = m_X/2m_a$ is the axion boost factor. Together with~\eqref{eq:axionLifetime} this yields the requirement
\begin{align}
    g_{a\gamma\gamma} \lesssim 10^{-6}\,
    \GeV^{-1}\, \left(\frac{\eV}{m_a} \right)^2  \left(\frac{10^{26}\,{\rm s}}{\tau_{\rm min}}\right)^{1/2}\left(\frac{m_X}{\GeV}\right)^{1/2} ,
\end{align}
where we have normalized to typical values. From this we conclude that for sub-eV axion masses, the indirect detection constraints on DM decay based on X-ray and $\gamma$-ray observations are avoided. In what follows we shall therefore assume $m_a\lesssim 1~\eV$.

The lightness of the ALPs as imposed by the indirect detection constraints implies that stellar cooling constraints on $g_{a\gamma\gamma}$ are in general applicable, as the axion is kinematically accessible in the interiors of stars. The leading constraint over a large mass range comes from globular cluster observations of the ratio of Horizontal Branch to Red Giant stars~\cite{Ayala:2014pea,Dolan:2022kul} which limits the axion-photon coupling to better than $g_{a\gamma\gamma}\lesssim  10^{-10}\,{\rm GeV}$. For even smaller axion masses, $m_a\lesssim 10^{-7}\,\eV$, a multitude of additional constraints derived from high energy astrophysics enter, further strengthening the bounds down to the $ 10^{-12}\,{\rm GeV}$ level (see~\cite{githublimits} for a compilation of various constraints). It is worthwhile noting that recent studies~\cite{DeRocco:2020xdt, Bar:2019ifz} have demonstrated various mechanisms by which the astrophysical bounds may not be robust, and can be alleviated/evaded. From this perspective, independent probes for ALPs with a well-controlled lab environment, such as proposed in this work (as well as other existing terrestrial searches), are highly desirable and can provide complementary information.

Finally, we note that even in the absence of a tree-level axion-electron coupling, an effective  $g_{aee}^{\rm eff}$ interaction can be induced at the loop-level. For $m_a \ll m_e$ Ref.~\cite{Bauer:2017ris} finds
\begin{align}
\label{eq:gaeeff}
g_{aee}^{\rm eff} = \dfrac{3\alpha m_e}{4\pi} g_{a\gamma\gamma} \left(\ln\dfrac{\Lambda^2}{m_e^2} - \dfrac{4}{3} \right) + \dots \,,
\end{align}
where $\Lambda$ is a UV scale and the dots represent additional contributions that may enter from the electroweak sector. Such radiatively induced coupling is subject to the severe bound from Red Giant stars, $g_{aee}\lesssim 10^{-13}$, see~\cite{Capozzi:2020cbu} and references therein.
Using~\eqref{eq:gaeeff} this stellar bound is then compatible with $g_{a\gamma\gamma} \lesssim 10^{-7} \,{\rm GeV}^{-1}$.

Once the ALPs reach a terrestrial detector, they can be detected via absorption through the IP process\footnote{Since in this work we will consider ALP scenarios with a long lifetime, the event rate from $a\rightarrow \gamma\gamma$ inside any detector is negligible.}. The differential cross section reads
\begin{align}
\dfrac{d\sigma_{\rm IP}}{d\Omega_a} = \dfrac{g_{a\gamma\gamma}^2 \alpha}{4\pi} \dfrac{p_a^4}{q^4} \sin^2 \theta_a F^2 (q)\,,
\label{eq:inverse_Primakoff_diffxsec}
\end{align}
where $p_a = |\vec{p}_a|$, $\theta_a$ is the scattering angle, $q^2 = m_a^2 - 2 E_\gamma (E_a - p_a \cos\theta_a)$ and $F (q)$ is the atomic/nuclear form factor, $E_\gamma$ is the energy of the outgoing photon. See Appendix~\ref{app:xsec_IP} and~\cite{Brdar:2020dpr} for a derivation. By examining the kinematics of the final states, we found that the outgoing photon is highly monochromatic, with $E_\gamma\approx E_a$, which constitutes a characteristic signal for the searches. 

\section{Experimental probes}
\label{sec:experiments}
We now study the prospects of constraining ALP DR with large volume neutrino detectors. 
We will particularly focus on a few benchmark experiments: Super-Kamiokande (SK), Hyper-Kamiokande (HK), Icecube-Deepcore, DUNE and JUNO. In water-based Cherenkov detectors such as SK/HK and IceCube, Cherenkov rings induced by photons and electrons are similar, thus the IP signal indiscriminately reveals itself as an excess in electron scattering-like events. On the other hand, DUNE and JUNO have the capacity of high efficiency photon identification, allowing us to capture the monochromatic photon signal without confusing it with electron scatterings. Due to the inability of distinguishing electron from photon, with SK and HK we expect a notable background for our signal from $e^-$ events due to diffuse supernova neutrinos for  $\mathcal{O}(10\text{--}30\,{\rm ~MeV})$ and atmospheric neutrinos for $\mathcal{O}(30\,~{\rm MeV}\text{--}1~ {\rm GeV})$. 
On the other hand, the monochromatic energetic single photon signal at DUNE and JUNO can be treated as background-free. We review the relevant details about these experiments in the following, along with the parameter choices made for our analysis.

\subsection{Super-Kamiokande}
\label{sec:exp_SK}
With the data accumulated, SK has the potential to constrain the parameter space of the scenario under consideration.
The fiducial mass of the SK detector is taken to be $22.5\,{\rm kton}$ and we take the following two energy bins for which backgrounds were reported \footnote{We note that there are also bins with higher $E_\gamma$ in~\cite{Super-Kamiokande:2017dch}. 
However, the number of observed events there is below the Monte-Carlo based background estimate. A detailed fit to the data is thus required to consistently set constraint on the axion parameter space, which is beyond the scope of this work.}. 
First, we consider the range of $E_\gamma = [16,88]\,{\rm MeV}$, utilizing the 1497 days of electron-recoil data from SK-I run in search for a diffuse supernova neutrino background~\cite{Super-Kamiokande:2011lwo}.
Within the signal region of  Cherenkov angle, a total of 239 events were observed with 238 background events expected from the best-fit background model; the background event rate is  $58.3/{\rm yr}$.
The detection efficiency is found in~\cite{Super-Kamiokande:2011lwo}. 
Another energy range we consider is $E_\gamma = [0.1,1.33]\,{\rm GeV}$, constrained by the lowest energy interval from an analysis of fully-contained single electron-like events in atmospheric neutrino search with SK-IV~\cite{Super-Kamiokande:2017dch}.
The exposure is $161.9\,$kton-yr and 4042 events after cuts were reported with an efficiency $\epsilon (E_\gamma = 0.5\,{\rm GeV}) = 0.93$~\cite{Super-Kamiokande:2017dch}. The number of background events in the same energy interval is estimated to be $3993$~\cite{Super-Kamiokande:2017dch}, equivalent to a background event rate of $554.9/{\rm yr}$.
We derive the $90\%$ C.L. limits on $(m_X, g_{a\gamma\gamma})$ parameter space using the following criterion~\cite{ParticleDataGroup:2020ssz} 
\begin{align}
\label{eq:bound_criteria}
 N_{\rm sig}^{\rm ALP} \leq {\rm Max}[0,N_{\rm obs} - N_{\rm bkg}] + 1.28 \sqrt{N_{\rm obs}}\,,
\end{align}
where $N_{\rm sig}^{\rm ALP}$, $N_{\rm obs}$, $N_{\rm bkg}$ are number of events induced by the ALPs, total observed events and background events, respectively.

\subsection{Hyper-Kamiokande}

As an upgraded version of SK, HK received the final approval for construction in 2019 and data-taking is expected to start in 2027.
The fiducial volume of the HK detector is about 25 times that of SK~\cite{Abe:2011ts,Hyper-KamiokandeWorkingGroup:2013hcb}.
We derive the projection of sensitivity based on the same energy intervals considered for SK,
assuming the same data-taking time as SK (1497 days for $E_\gamma = [16,88]\,{\rm MeV}$ and 2626 days for $E_\gamma = [0.1,1.33]\,{\rm GeV}$). We rescale the background events for SK according to the difference in fiducial volume in order to estimate the background events for HK (\textit{i.e.}, the background event rates are $1457.5/{\rm yr}$ for $E_\gamma = [16,88]\,{\rm MeV}$ and $13872.5/{\rm yr}$ for $E_\gamma = [0.1,1.33]\,{\rm GeV}$).
With a conservative detection efficiency of 0.8, the projected sensitivity of HK can be derived by requiring $N_{\rm sig}^{\rm ALP} \leq 1.28 \sqrt{N_{\rm bkg}}$ assuming $N_{\rm obs} = N_{\rm bkg}$. 

\subsection{IceCube}
Cherenkov detectors such as
IceCube has the advantage of larger volume compared to SK/HK. Nevertheless, this advantage is counteracted by their higher energy detection thresholds. Primary sensitivity is therefore expected in the higher progenitor mass region.
Here we focus on IceCube, which has the largest volume, $1\,$Gton mass with $100\,$GeV energy detection threshold, while its DeepCore sub-array has a 10~Mton mass and $\mathcal{O}(10)\,$GeV threshold.
Like with SK and HK, the boosted ALP signal would reveal itself as electron-like events, and the existing atmospheric $\nu_e$ data~\cite{IceCube:2015mgt} leads to potential constraints.

We consider the the measurement of the atmospheric $\nu_e$ flux in the energy range between $80\,{\rm GeV}$ and $6\,{\rm TeV}$ from DeepCore~\cite{IceCube:2012jwm}.
With 281 live days, a total of 1029 events are observed, while the total number of predicted events is 1007~\cite{Honda:2006qj}, which we take as the background for the ALP search.
The detection efficiency is derived by dividing the effective volume reported in~\cite{IceCube:2012jwm} by the fiducial volume of DeepCore.
In addition, we consider data from IceCube with an energy threshold of $100\,{\rm GeV}$~\cite{IceCube:2015mgt}. 
A total of 1078 fully contained events are observed within 332 live days and with energy below $100\,{\rm TeV}$. The estimated number of background events based on MC simulation is $N_{\rm bkg} = 1071$. The detection efficiency is deduced from the fiducial volume of $1\,{\rm Gton}$ and an effective volume for $\nu_e$ level 4 shown in~\cite{IceCube:2015mgt}. 
For both DeepCore and IceCube, the constraints on ALP parameter space are derived based on Eq.~\eqref{eq:bound_criteria}.

Recently, a major extension, the IceCube Upgrade~\cite{Ishihara:2019aao}, was approved and will launch in mid 2020s, which has an effective volume of 2 Mton and lowered detection threshold of 1 GeV. In lack of a reliable background estimate as of now, we defer the investigation of the potential with IceCube Upgrade for future work.

\subsection{DUNE}
The DUNE is currently under construction and expected to start taking data around 2028 (far detectors). 
The DUNE far detectors (DUNE-FD) are of two types:
a single phase LArTPC module (DUNE-SP)~\cite{DUNE:2020txw} and a dual phase module (DUNE-DP)~\cite{DUNE:2018mlo}. 
DUNE-SP is a 40-kilo-tonne Single Phase LArTPC, which can search for MeV to GeV-scale neutrino interactions.
It has 6000 photon-detection system (PDS) channels with minimal detection threshold of 10 MeV~\cite{DUNE:2020ypp}. DUNE-DP~\cite{DUNE:2018mlo} has a gas phase in addition to the LArTPC, which has the ionization signal obtained in the gas phase, as an attempt to achieve a tunable signal over background rate and a lower detection threshold. It has a 12 kton active mass of LArTPC.

In this work, we consider a DUNE-SP-like detector and study the sensitivity reach with 40 kton active LAr mass; the sensitivity of DUNE-DP is harder to estimate due to its more complicated experimental setup.
We adopt a detection threshold of $10\,{\rm MeV}$ (this threshold may be pushed lower in reality) and follow~\cite{DUNE:2020ypp} for the reconstruction efficiency of the photon signal. 
An advantage of DUNE-like detectors over the Cherenkov detectors (\textit{e.g.} SK/HK) is that, the photon signal from ALP absorption can be better distinguished from electron events based on dedicated identification. Thus the potential background here is dominated by genuine photons from the decay of $\pi^0$'s that are produced from hadronic interaction of the incident atmospheric/DSN neutrinos. As discussed in~\cite{Brdar:2020dpr} in the context of beam-produced ALPs, such a $\gamma$ background can be effectively mitigated by vetoing on additional hadronic activities and the second shower (typical for $\pi^0\rightarrow \gamma\gamma$). In addition, the characteristic monochromatic single photon signal should lend itself to further discriminatory power from the kinematic information. We reserve a quantitative, dedicated analysis in this regard for future work. In this work we will assume zero background at DUNE.
Therefore, we derive the sensitivity reach based on the number of signal events $N_{\rm sig}^{\rm ALP}$ per year. We take $N_{\rm sig}^{\rm ALP}/{\rm yr} = 1$ and $3$ as two benchmarks, representing the aggressive and conservative projections.

\subsection{JUNO}
The JUNO  neutrino detector is currently under construction in China with data-taking expected to begin in 2023.
JUNO will be equipped with the to-date largest liquid scintillator detector with a fiducial volume of around 20\,kton, surrounded by around 53000 photomultipliers.
We set the detection threshold to $12\,{\rm MeV}$~\cite{JUNO:2022lpc}. 
An optimistic detection efficiency of monochromatic-$\gamma$ is assumed to be unity for simplicity, given its high performance in particle discrimination~\cite{Rebber:2020xfi} and high photon detection efficiency~\cite{JUNO:2021vlw}.  
With the considerations similar to that of DUNE, we consider the background to be well-controlled and approximately zero for the monochromatic-$\gamma$ signals.
Therefore, we show the projected sensitivity based on $N_{\rm sig}^{\rm ALP}$ per year, for which the conservative (aggressive) one corresponds to $N_{\rm sig}^{\rm ALP}/{\rm yr} = 3 \,(1)$. 

\section{Analysis results, experimental sensitivities}
\label{sec:results}

\begin{figure}[t]
\centering
\includegraphics*[width=0.98\columnwidth]{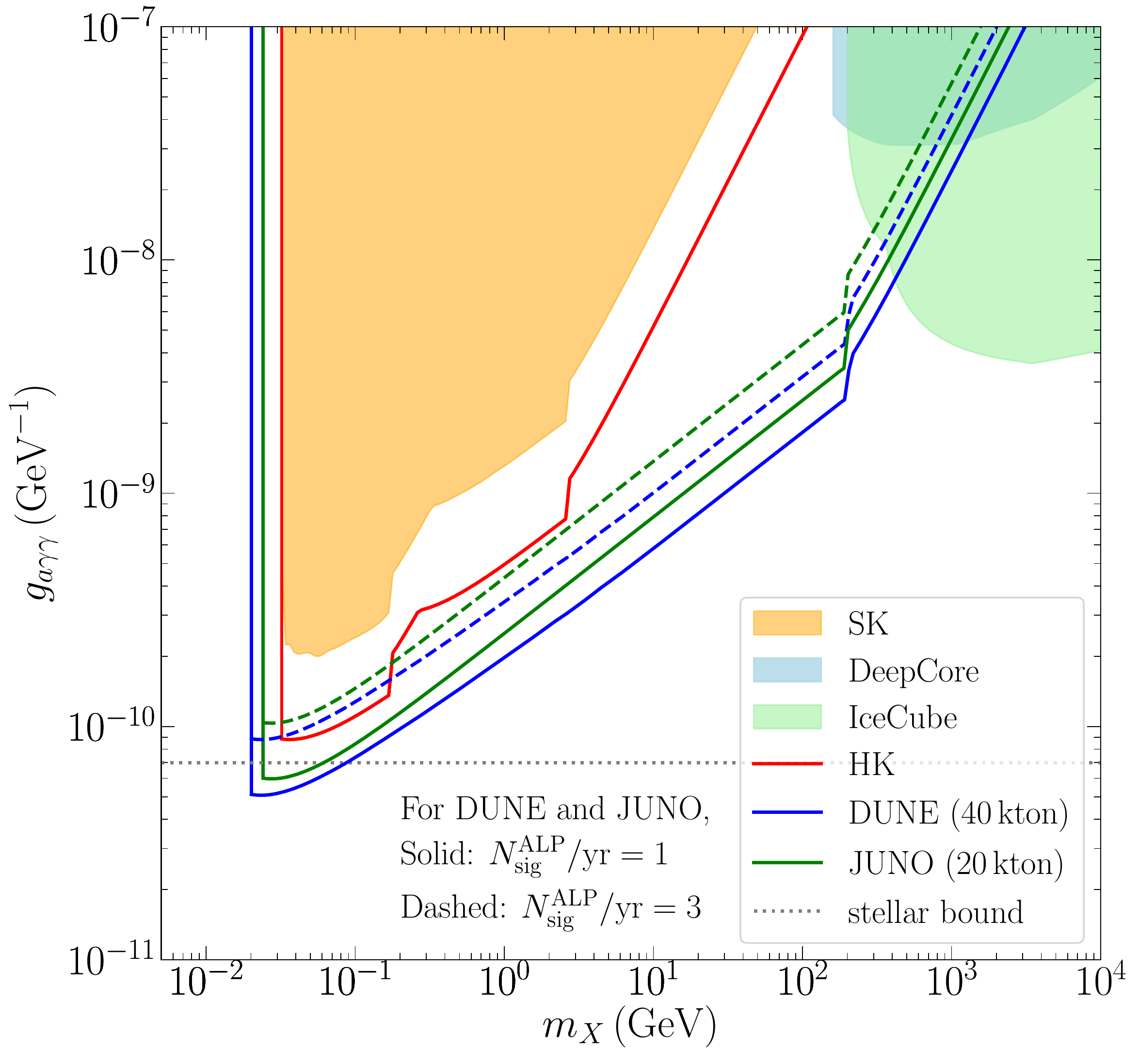}
\caption{ Current bounds (shaded region) and sensitivity reach (curves) on $(m_X, g_{a\gamma\gamma})$ of large volume neutrino experiments considered in this work. Stability reasons require $m_a\lesssim \eV$ and the dashed horizontal line shows the stellar bound on $g_{a\gamma\gamma}$. In making the plot, we assumed a progenitor lifetime of the  $\tau_ X = 35\,t_0$ compatible with current constraints on DM decay. Since the mono-$\gamma$ signal is largely background-free in DUNE and JUNO, we show the conservative (3 events per year, dashed curves) and aggressive (1 event per year, solid curves) projections for demonstration.
We truncate at $m_X = 10^4\,{\rm GeV}$ as the sensitivity reach of experiments only becomes weaker due to a smaller flux. 
We find that both a DUNE-like experiment and JUNO can improve the sensitivity for $m_X \sim \mathcal{O} (10\text{--}100\,{\rm GeV}$) beyond astrophysical constraints.
}
\label{fig:sensitivity_reach}
\end{figure}

We now present the sensitivity reach of the experiments mentioned above. In Fig.~\ref{fig:sensitivity_reach} the shaded regions correspond to the current sensitivity from SK and DeepCore/IceCube, while the curves represent the projections of future sensitivity for HK, DUNE, and JUNO. The signal event rate is obtained by integrating over the energy range of $[E_\gamma^{\rm min}, E_\gamma^{\rm max}]$ for the experiment in consideration. $E_\gamma^{\rm min}$ is the detection threshold. For SK/HK and Icecube, $E_\gamma^{\rm max}$ indicates the upper bounds of the particular energy bins we used for background estimates based on existing data, as specified earlier (two bins for SK/HK). With DUNE and JUNO, there is no detailed binning information for the searches available now, and we expect them to be approximately background free for the ALP photon signal, thus for them $E_\gamma^{\rm max}$ may be taken as the maximal energy that can be contained within the detector. We make a conservative choice of $E_\gamma^{\rm max}=100$ GeV for DUNE and JUNO. The constraint/reach at low $m_X$ is only sensitive to $E_\gamma^{\rm min}$, while larger $E_\gamma^{\rm max}$ would improve the sensitivity at larger $m_X$ (typically superseded by the stellar bound). The galactic component dominates the flux at low $m_X < 2 E_\gamma^{\rm max}$.
For $m_X $ larger than $2 E_\gamma^{\rm max}$, sensitivity is lost as the constraint is then necessarily driven by a smaller, redshifted fraction of the extragalactic flux. 

~~Among the considered experiments, both the DUNE-like detector and JUNO are capable of probing $g_{a\gamma\gamma}$ values below current astrophysical sensitivity.
On the other hand, for the Cherenkov detectors SK and HK, the mono-$\gamma$ signal is subject to background, which precludes a competitive sensitivity with respect to  DUNE and JUNO.

As can be seen from Fig.~\ref{fig:sensitivity_reach}, SK already provides effective constraints in the lower mass range with best sensitivity at $m_X\sim 50~\MeV$, while HK will be able to probe $g_{a\gamma\gamma}$ close to stellar bound at a value $10^{-10}\,\GeV^{-1}$. Thanks to a lower energy threshold, the  DUNE-like detector and JUNO will be able to reach beyond the astrophysical limits with the best sensitivity of $g_{a\gamma\gamma}\sim 5\times 10^{-11}\,\GeV^{-1}$ in the sub-GeV progenitor mass region.  Finally, we discover that the sensitivity of IceCube is relatively weak, well above  $10^{-9}\,\GeV^{-1}$  and well in the astrophysically disfavored region.

\section{Conclusions}
\label{sec:conclusions}
In this work we propose a new search avenue for cosmogenic relativistic axions with neutrino detectors. We considered the benchmark example of axion production from dark matter decay and the characteristic monochromatic single photon signal for detection induced by the inverse-Primakoff effect. Other ramifications are worth further investigation, including other cosmic sources for such energetic axions and other axion interaction modes. In addition, we are not being exhaustive about relevant neutrino experiments, and the potential of other facilities such as ANTARES, KM3NeT, and KamLAND-Zen \cite{ANTARES:2021cwc, KM3Net:2016zxf,Shirai:2017jyz} are worth exploring.
This proposal well complements existing axion searches mostly based on cold relics, while it provides new targets for dark matter related BSM physics searches with future neutrino experiments including DUNE, HK and JUNO.

\section*{Acknowledgement}
We thank Yingying Li and Yun-Tse Tsai for valuable discussions. YC is supported in part by the U.S. Department of Energy under award number DE-SC0008541. JLK and YDT are supported by the U.S. National Science Foundation (NSF) Theoretical Physics Program, Grant PHY-1915005. Part of this manuscript has been authored by Fermi Research Alliance, LLC under Contract No. DE-AC02-07CH11359 with the U.S. Department of Energy, Office of Science, Office of High Energy Physics.
YDT thanks the Kavli Institute for Theoretical Physics for hospitality, supported partly by the National Science Foundation under Grant No. NSF PHY-1748958, and the Aspen Center for Physics, funded partially by the National Science Foundation grant PHY-1607611, for their hospitality.

\appendix 
\section{Flux of ALPs from dark matter decay}
\begin{figure}[t]
 \centering
 \includegraphics*[width=0.98\columnwidth]{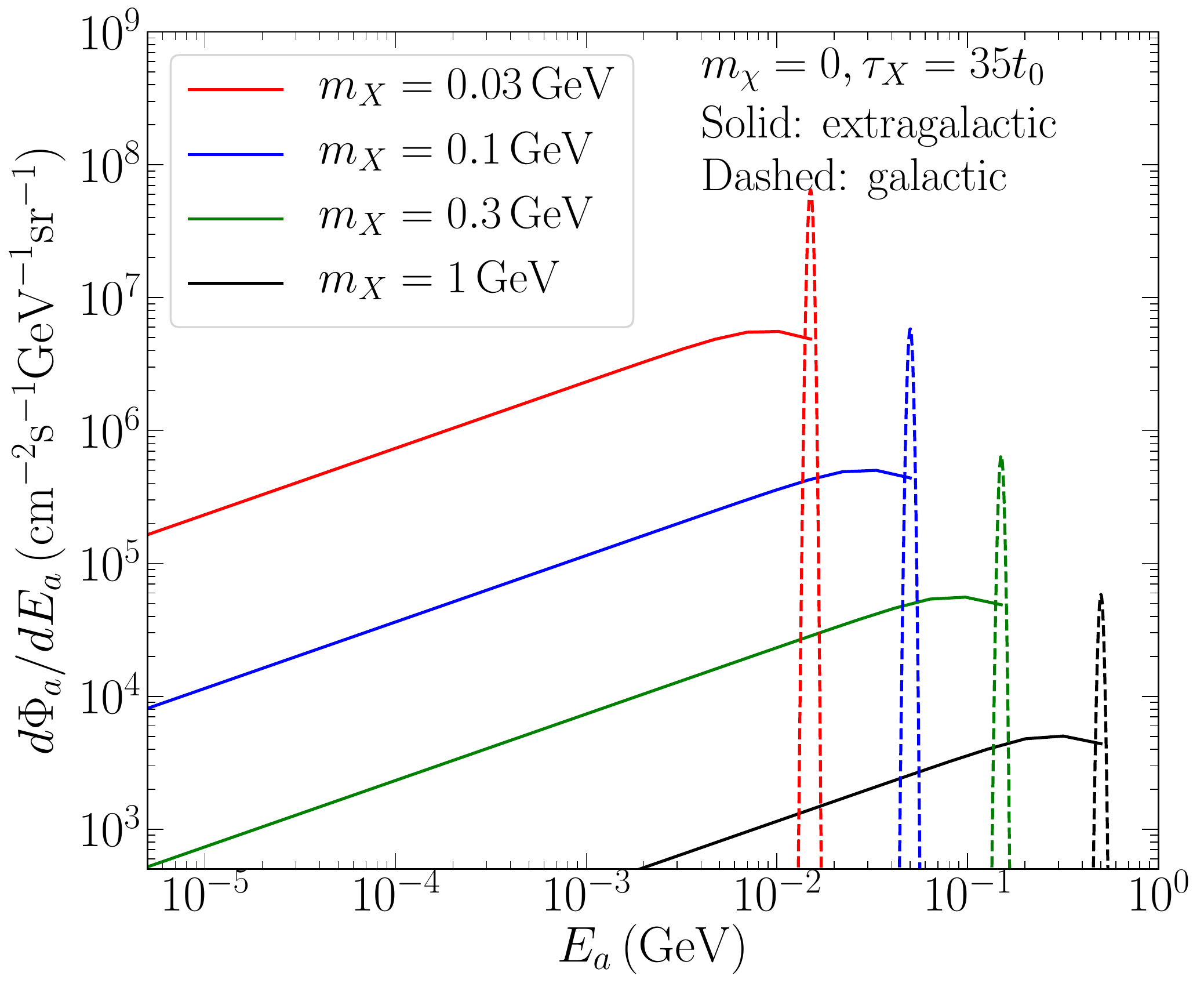}
 \caption{The energy-differential flux of the ALPs from DM decay. Different colors represent different progenitor masses.}
 \label{fig:axion_flux}
\end{figure}

In Fig.~\ref{fig:axion_flux} we illustrate the energy spectra of the flux of the ALPs from DM decay, including both galactic and extragalactic components. For better demonstration of the monochromatic galactic flux, we apply a $2\%$ Gaussian smearing to it.

\section{ALP-electron interaction}

The effective Lagrangian for interaction between ALP and electron is given by
\begin{align}
\mathcal{L}_e = \dfrac{g_{aee}}{2m_e} (\partial_\mu a) \bar{e} \gamma^\mu \gamma^5 e\,,
\end{align}
where $g_{aee}$ is a dimensionless coupling and $m_e$ is electron mass. The one-loop diagram that induces such an effective axion-electron coupling through $g_{a\gamma\gamma}$ yields in magnitude~\cite{Bauer:2017ris}
\begin{align}
g_{a\gamma\gamma}^{\rm eff} = \dfrac{\alpha}{\pi m_e} g_{aee} B_1(4m_e^2 /m_a^2)\,,
\end{align}
where $B_1(x) = 1- x \left[\sin^{-1}(1/\sqrt{x}) \right]^2$.

\section{Cross section of inverse-Primakoff process}
\label{app:xsec_IP}

For $a (p_1)+ N (p_2) \rightarrow \gamma (p_3) + N(p_4)$, the lab-frame kinematics where the nucleus is at rest is given by
\begin{align}
 &p_1 = (E_a,0,0, p_a)\,, \quad p_2= (m_N, 0,0,0)\,, \nonumber \\
 &p_3 = (E_\gamma, 0 , E_\gamma \sin\theta_\gamma, E_\gamma \cos\theta_\gamma)\,, \quad p_4 = p_1 + p_2 - p_3\,,
\end{align}
where $m_N$ is nucleus mass and $E_a = \sqrt{p_a^2 + m_a^2}$.
Assuming the scattering is elastic, \textit{i.e.}, $p_4^2 = m_N^2$, we  obtain the relation
\begin{align}
E_\gamma = \dfrac{-2E_a m_N -m_a^2}{2 (\cos\theta_\gamma p_a -  E_a -  m_N)}\,.
\end{align}
The scattering amplitude reads
\begin{align}
\label{eq:amp_Primakoff}
 \mathcal{M}_{\rm IP} = \dfrac{g_{\mu\nu}}{t^2} j_a^\mu j_N^\nu\,,
\end{align}
where $t\equiv (p_1 - p_3)^2 \equiv q^2$, $j_a^\mu = g_{a\gamma\gamma} \epsilon^{\delta\mu\alpha\beta} \varepsilon_\delta(p_3) p_{3,\alpha} q_\beta$, and $j_N^\nu = e F(q) (p_2^\nu + p_4^\nu)$ taking heavy scalar target limit with $F(q)$ being the atomic/nuclear form factor. Squaring Eq.~\eqref{eq:amp_Primakoff} and summing over final photon polarization, we obtain the Lorentz-invariant squared amplitude for the IP process 
\begin{align}
\left|\mathcal{M}_{\rm IP} \right|^2 = -\dfrac{\pi\alpha F^2(q) g_{a\gamma\gamma}^2 }{t^2} &\left\lbrace m_a^4 (4m_N^2 +3 t) -2m_a^2 t (4s +t) \right. \nonumber \\
& \left. + 4t\left[(m_N^2 -s)^2 +st \right]\right\rbrace\,,
\end{align}
where $s \equiv (p_1 + p_2)^2$. In the lab frame, we write for $t$ and $s$,
\begin{align}
   &t = q^2 = m_a^2 - 2E_\gamma(E_a - p_a \cos\theta_\gamma)\,, \nonumber \\ 
   &s= m_N ( 2 E_\gamma + m_N)\,.
\end{align}
The general form of differential cross section reads
\begin{align}
  \dfrac{d\sigma_{\rm IP}}{d\Omega_a} = \dfrac{1}{64\pi^2 s} \left|\mathcal{M}_{\rm IP} \right|^2\,,
\end{align}
which reduces to Eq.~\eqref{eq:inverse_Primakoff_diffxsec} if we consider a light ALP and small momentum transfer relative to nucleus mass, \textit{i.e.},  $m_a, p_a \ll m_N$. The Primakoff process and its inverse are related through  $d\sigma_{\rm P}/d\Omega_a = (d\sigma_{\rm IP}/d\Omega_a)/2$ from averaging over the polarizations of initial photon.

\begin{figure}[t]
\centering
\includegraphics*[width=0.95\columnwidth]{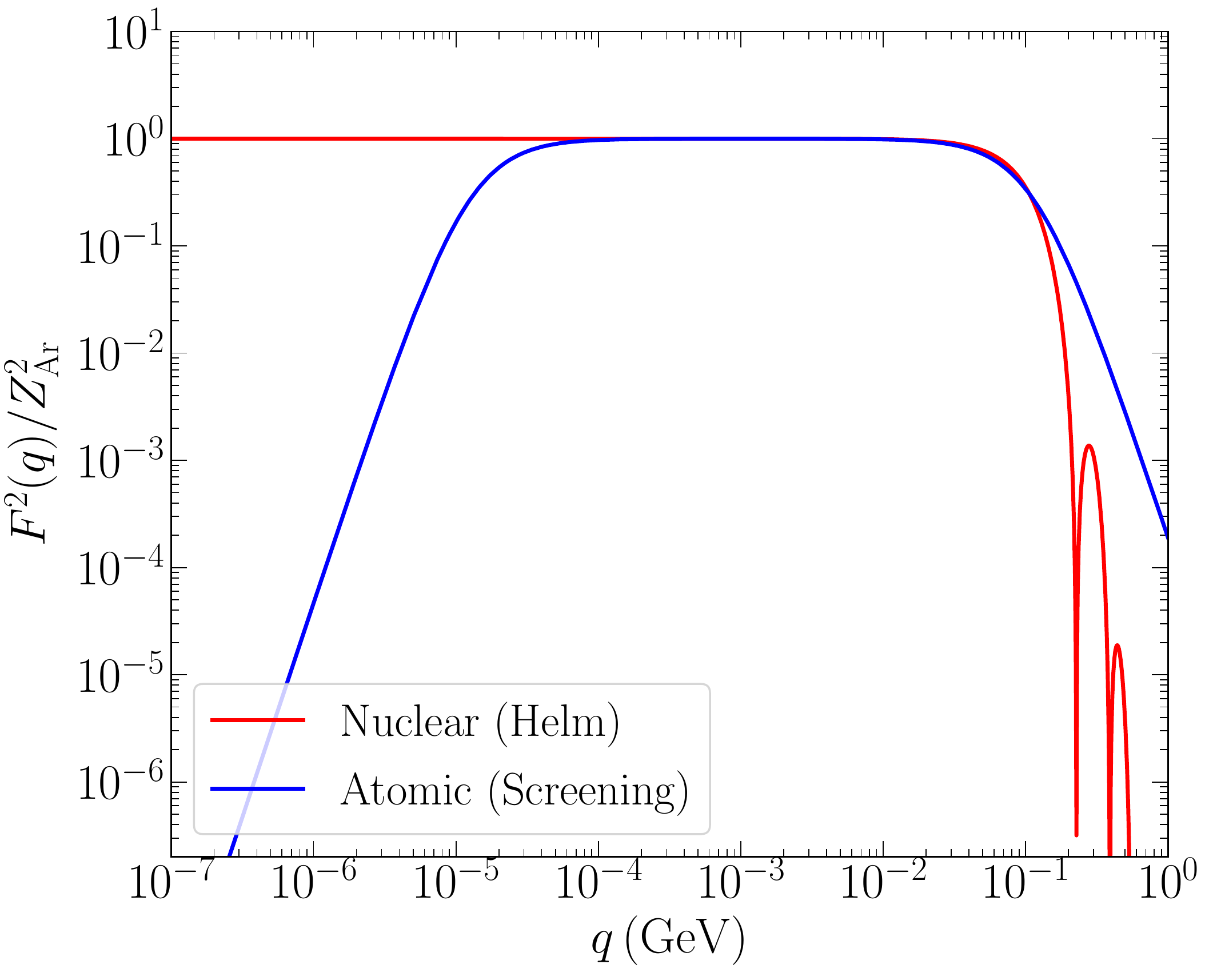}
\caption{Values of atomic and nuclear form factors as a function of momentum transfer $q$ for Argon.
}
\label{fig:form_factor}
\end{figure}

The correct form factor $F(q)$ to be used depends on the momentum transfer $q$.
For $q \ll {\rm MeV}$, we  adopt the atomic form factor which includes electron screening of the nuclear charge~\cite{Kim:1973he,Tsai:1973py}:  
\begin{align}
\label{eq:screening_ff}
  F^2(q) = Z^2 \left[ \dfrac{a^2(Z) q^2}{1+a^2(Z) q^2} \dfrac{1}{1+q^2/d(A)}\right]^2\,,
\end{align}
where $a(Z) = 111Z^{-1/3}/ m_e$ and $d(A) = 0.164\,{\rm GeV}^2 A^{-2/3}$ with $Z$ and $A$ being atomic and mass number of the nucleus. 
When $q$ is large enough to probe the nucleus, we use the the Helm form factor~\cite{Helm:1956zz}
\begin{align}
\label{eq:helm_ff}
F^2(q) = Z^2 \left[ \dfrac{3j_1(q R_1)}{q R_1}\right]^2 e^{-q^2 s^2}\,,
\end{align}
where $R_1 = \sqrt{(1.23 A^{1/3} -0.6)^2 + 2.18} \,{\rm fm}$, $j_1 $ is the spherical Bessel function, and $s = 0.9\,{\rm fm}$.
The comparison of Eq.~\eqref{eq:screening_ff} and Eq.~\eqref{eq:helm_ff} for Argon is given in Fig.~\ref{fig:form_factor}. For $q \ll {\rm MeV}$, we can see the suppression due to electron screening in atomic form factor, while for $q \gg {\rm MeV}$ nuclear form factor accounts  for the resolved nuclear structure.

\bibliography{refs_AxionDUNE}

\begin{thebibliography}{73}%
\makeatletter
\providecommand \@ifxundefined [1]{%
 \@ifx{#1\undefined}
}%
\providecommand \@ifnum [1]{%
 \ifnum #1\expandafter \@firstoftwo
 \else \expandafter \@secondoftwo
 \fi
}%
\providecommand \@ifx [1]{%
 \ifx #1\expandafter \@firstoftwo
 \else \expandafter \@secondoftwo
 \fi
}%
\providecommand \natexlab [1]{#1}%
\providecommand \enquote  [1]{``#1''}%
\providecommand \bibnamefont  [1]{#1}%
\providecommand \bibfnamefont [1]{#1}%
\providecommand \citenamefont [1]{#1}%
\providecommand \href@noop [0]{\@secondoftwo}%
\providecommand \href [0]{\begingroup \@sanitize@url \@href}%
\providecommand \@href[1]{\@@startlink{#1}\@@href}%
\providecommand \@@href[1]{\endgroup#1\@@endlink}%
\providecommand \@sanitize@url [0]{\catcode `\\12\catcode `\$12\catcode
  `\&12\catcode `\#12\catcode `\^12\catcode `\_12\catcode `\%12\relax}%
\providecommand \@@startlink[1]{}%
\providecommand \@@endlink[0]{}%
\providecommand \url  [0]{\begingroup\@sanitize@url \@url }%
\providecommand \@url [1]{\endgroup\@href {#1}{\urlprefix }}%
\providecommand \urlprefix  [0]{URL }%
\providecommand \Eprint [0]{\href }%
\providecommand \doibase [0]{http://dx.doi.org/}%
\providecommand \selectlanguage [0]{\@gobble}%
\providecommand \bibinfo  [0]{\@secondoftwo}%
\providecommand \bibfield  [0]{\@secondoftwo}%
\providecommand \translation [1]{[#1]}%
\providecommand \BibitemOpen [0]{}%
\providecommand \bibitemStop [0]{}%
\providecommand \bibitemNoStop [0]{.\EOS\space}%
\providecommand \EOS [0]{\spacefactor3000\relax}%
\providecommand \BibitemShut  [1]{\csname bibitem#1\endcsname}%
\let\auto@bib@innerbib\@empty
\bibitem [{\citenamefont {Preskill}\ \emph {et~al.}(1983)\citenamefont
  {Preskill}, \citenamefont {Wise},\ and\ \citenamefont
  {Wilczek}}]{Preskill:1982cy}%
  \BibitemOpen
  \bibfield  {author} {\bibinfo {author} {\bibfnamefont {J.}~\bibnamefont
  {Preskill}}, \bibinfo {author} {\bibfnamefont {M.~B.}\ \bibnamefont {Wise}},
  \ and\ \bibinfo {author} {\bibfnamefont {F.}~\bibnamefont {Wilczek}},\ }\href
  {\doibase 10.1016/0370-2693(83)90637-8} {\bibfield  {journal} {\bibinfo
  {journal} {Phys. Lett. B}\ }\textbf {\bibinfo {volume} {120}},\ \bibinfo
  {pages} {127} (\bibinfo {year} {1983})}\BibitemShut {NoStop}%
\bibitem [{\citenamefont {Abbott}\ and\ \citenamefont
  {Sikivie}(1983)}]{Abbott:1982af}%
  \BibitemOpen
  \bibfield  {author} {\bibinfo {author} {\bibfnamefont {L.~F.}\ \bibnamefont
  {Abbott}}\ and\ \bibinfo {author} {\bibfnamefont {P.}~\bibnamefont
  {Sikivie}},\ }\href {\doibase 10.1016/0370-2693(83)90638-X} {\bibfield
  {journal} {\bibinfo  {journal} {Phys. Lett. B}\ }\textbf {\bibinfo {volume}
  {120}},\ \bibinfo {pages} {133} (\bibinfo {year} {1983})}\BibitemShut
  {NoStop}%
\bibitem [{\citenamefont {Dine}\ and\ \citenamefont
  {Fischler}(1983)}]{Dine:1982ah}%
  \BibitemOpen
  \bibfield  {author} {\bibinfo {author} {\bibfnamefont {M.}~\bibnamefont
  {Dine}}\ and\ \bibinfo {author} {\bibfnamefont {W.}~\bibnamefont
  {Fischler}},\ }\href {\doibase 10.1016/0370-2693(83)90639-1} {\bibfield
  {journal} {\bibinfo  {journal} {Phys. Lett. B}\ }\textbf {\bibinfo {volume}
  {120}},\ \bibinfo {pages} {137} (\bibinfo {year} {1983})}\BibitemShut
  {NoStop}%
\bibitem [{\citenamefont {Dror}\ \emph {et~al.}(2021)\citenamefont {Dror},
  \citenamefont {Murayama},\ and\ \citenamefont {Rodd}}]{Dror:2021nyr}%
  \BibitemOpen
  \bibfield  {author} {\bibinfo {author} {\bibfnamefont {J.~A.}\ \bibnamefont
  {Dror}}, \bibinfo {author} {\bibfnamefont {H.}~\bibnamefont {Murayama}}, \
  and\ \bibinfo {author} {\bibfnamefont {N.~L.}\ \bibnamefont {Rodd}},\ }\href
  {\doibase 10.1103/PhysRevD.103.115004} {\bibfield  {journal} {\bibinfo
  {journal} {Phys. Rev. D}\ }\textbf {\bibinfo {volume} {103}},\ \bibinfo
  {pages} {115004} (\bibinfo {year} {2021})},\ \Eprint
  {http://arxiv.org/abs/2101.09287} {arXiv:2101.09287 [hep-ph]} \BibitemShut
  {NoStop}%
\bibitem [{\citenamefont {Pospelov}\ \emph {et~al.}(2008)\citenamefont
  {Pospelov}, \citenamefont {Ritz},\ and\ \citenamefont
  {Voloshin}}]{Pospelov:2008jk}%
  \BibitemOpen
  \bibfield  {author} {\bibinfo {author} {\bibfnamefont {M.}~\bibnamefont
  {Pospelov}}, \bibinfo {author} {\bibfnamefont {A.}~\bibnamefont {Ritz}}, \
  and\ \bibinfo {author} {\bibfnamefont {M.~B.}\ \bibnamefont {Voloshin}},\
  }\href {\doibase 10.1103/PhysRevD.78.115012} {\bibfield  {journal} {\bibinfo
  {journal} {Phys. Rev. D}\ }\textbf {\bibinfo {volume} {78}},\ \bibinfo
  {pages} {115012} (\bibinfo {year} {2008})},\ \Eprint
  {http://arxiv.org/abs/0807.3279} {arXiv:0807.3279 [hep-ph]} \BibitemShut
  {NoStop}%
\bibitem [{\citenamefont {An}\ \emph {et~al.}(2015)\citenamefont {An},
  \citenamefont {Pospelov}, \citenamefont {Pradler},\ and\ \citenamefont
  {Ritz}}]{An:2014twa}%
  \BibitemOpen
  \bibfield  {author} {\bibinfo {author} {\bibfnamefont {H.}~\bibnamefont
  {An}}, \bibinfo {author} {\bibfnamefont {M.}~\bibnamefont {Pospelov}},
  \bibinfo {author} {\bibfnamefont {J.}~\bibnamefont {Pradler}}, \ and\
  \bibinfo {author} {\bibfnamefont {A.}~\bibnamefont {Ritz}},\ }\href {\doibase
  10.1016/j.physletb.2015.06.018} {\bibfield  {journal} {\bibinfo  {journal}
  {Phys. Lett. B}\ }\textbf {\bibinfo {volume} {747}},\ \bibinfo {pages} {331}
  (\bibinfo {year} {2015})},\ \Eprint {http://arxiv.org/abs/1412.8378}
  {arXiv:1412.8378 [hep-ph]} \BibitemShut {NoStop}%
\bibitem [{\citenamefont {Bloch}\ \emph {et~al.}(2017)\citenamefont {Bloch},
  \citenamefont {Essig}, \citenamefont {Tobioka}, \citenamefont {Volansky},\
  and\ \citenamefont {Yu}}]{Bloch:2016sjj}%
  \BibitemOpen
  \bibfield  {author} {\bibinfo {author} {\bibfnamefont {I.~M.}\ \bibnamefont
  {Bloch}}, \bibinfo {author} {\bibfnamefont {R.}~\bibnamefont {Essig}},
  \bibinfo {author} {\bibfnamefont {K.}~\bibnamefont {Tobioka}}, \bibinfo
  {author} {\bibfnamefont {T.}~\bibnamefont {Volansky}}, \ and\ \bibinfo
  {author} {\bibfnamefont {T.-T.}\ \bibnamefont {Yu}},\ }\href {\doibase
  10.1007/JHEP06(2017)087} {\bibfield  {journal} {\bibinfo  {journal} {JHEP}\
  }\textbf {\bibinfo {volume} {06}},\ \bibinfo {pages} {087} (\bibinfo {year}
  {2017})},\ \Eprint {http://arxiv.org/abs/1608.02123} {arXiv:1608.02123
  [hep-ph]} \BibitemShut {NoStop}%
\bibitem [{\citenamefont {Brdar}\ \emph {et~al.}(2021)\citenamefont {Brdar},
  \citenamefont {Dutta}, \citenamefont {Jang}, \citenamefont {Kim},
  \citenamefont {Shoemaker}, \citenamefont {Tabrizi}, \citenamefont
  {Thompson},\ and\ \citenamefont {Yu}}]{Brdar:2020dpr}%
  \BibitemOpen
  \bibfield  {author} {\bibinfo {author} {\bibfnamefont {V.}~\bibnamefont
  {Brdar}}, \bibinfo {author} {\bibfnamefont {B.}~\bibnamefont {Dutta}},
  \bibinfo {author} {\bibfnamefont {W.}~\bibnamefont {Jang}}, \bibinfo {author}
  {\bibfnamefont {D.}~\bibnamefont {Kim}}, \bibinfo {author} {\bibfnamefont
  {I.~M.}\ \bibnamefont {Shoemaker}}, \bibinfo {author} {\bibfnamefont
  {Z.}~\bibnamefont {Tabrizi}}, \bibinfo {author} {\bibfnamefont
  {A.}~\bibnamefont {Thompson}}, \ and\ \bibinfo {author} {\bibfnamefont
  {J.}~\bibnamefont {Yu}},\ }\href {\doibase 10.1103/PhysRevLett.126.201801}
  {\bibfield  {journal} {\bibinfo  {journal} {Phys. Rev. Lett.}\ }\textbf
  {\bibinfo {volume} {126}},\ \bibinfo {pages} {201801} (\bibinfo {year}
  {2021})},\ \Eprint {http://arxiv.org/abs/2011.07054} {arXiv:2011.07054
  [hep-ph]} \BibitemShut {NoStop}%
\bibitem [{\citenamefont {Khatiwada}\ \emph {et~al.}(2021)\citenamefont
  {Khatiwada} \emph {et~al.}}]{ADMX:2020ote}%
  \BibitemOpen
  \bibfield  {author} {\bibinfo {author} {\bibfnamefont {R.}~\bibnamefont
  {Khatiwada}} \emph {et~al.} (\bibinfo {collaboration} {ADMX}),\ }\href
  {\doibase 10.1063/5.0037857} {\bibfield  {journal} {\bibinfo  {journal} {Rev.
  Sci. Instrum.}\ }\textbf {\bibinfo {volume} {92}},\ \bibinfo {pages} {124502}
  (\bibinfo {year} {2021})},\ \Eprint {http://arxiv.org/abs/2010.00169}
  {arXiv:2010.00169 [astro-ph.IM]} \BibitemShut {NoStop}%
\bibitem [{\citenamefont {Garcon}\ \emph {et~al.}(2017)\citenamefont {Garcon}
  \emph {et~al.}}]{Garcon:2017ixh}%
  \BibitemOpen
  \bibfield  {author} {\bibinfo {author} {\bibfnamefont {A.}~\bibnamefont
  {Garcon}} \emph {et~al.},\ }\href {\doibase 10.1088/2058-9565/aa9861} {\
  (\bibinfo {year} {2017}),\ 10.1088/2058-9565/aa9861},\ \Eprint
  {http://arxiv.org/abs/1707.05312} {arXiv:1707.05312 [physics.ins-det]}
  \BibitemShut {NoStop}%
\bibitem [{\citenamefont {Kahn}(2018)}]{Kahn:2018fgp}%
  \BibitemOpen
  \bibfield  {author} {\bibinfo {author} {\bibfnamefont {Y.}~\bibnamefont
  {Kahn}} (\bibinfo {collaboration} {ABRACADABRA}),\ }\href {\doibase
  10.1007/978-3-319-92726-8_16} {\bibfield  {journal} {\bibinfo  {journal}
  {Springer Proc. Phys.}\ }\textbf {\bibinfo {volume} {211}},\ \bibinfo {pages}
  {135} (\bibinfo {year} {2018})}\BibitemShut {NoStop}%
\bibitem [{\citenamefont {Anastassopoulos}\ \emph {et~al.}(2017)\citenamefont
  {Anastassopoulos} \emph {et~al.}}]{CAST:2017uph}%
  \BibitemOpen
  \bibfield  {author} {\bibinfo {author} {\bibfnamefont {V.}~\bibnamefont
  {Anastassopoulos}} \emph {et~al.} (\bibinfo {collaboration} {CAST}),\ }\href
  {\doibase 10.1038/nphys4109} {\bibfield  {journal} {\bibinfo  {journal}
  {Nature Phys.}\ }\textbf {\bibinfo {volume} {13}},\ \bibinfo {pages} {584}
  (\bibinfo {year} {2017})},\ \Eprint {http://arxiv.org/abs/1705.02290}
  {arXiv:1705.02290 [hep-ex]} \BibitemShut {NoStop}%
\bibitem [{\citenamefont {Armengaud}\ \emph {et~al.}(2019)\citenamefont
  {Armengaud} \emph {et~al.}}]{IAXO:2019mpb}%
  \BibitemOpen
  \bibfield  {author} {\bibinfo {author} {\bibfnamefont {E.}~\bibnamefont
  {Armengaud}} \emph {et~al.} (\bibinfo {collaboration} {IAXO}),\ }\href
  {\doibase 10.1088/1475-7516/2019/06/047} {\bibfield  {journal} {\bibinfo
  {journal} {JCAP}\ }\textbf {\bibinfo {volume} {06}},\ \bibinfo {pages} {047}
  (\bibinfo {year} {2019})},\ \Eprint {http://arxiv.org/abs/1904.09155}
  {arXiv:1904.09155 [hep-ph]} \BibitemShut {NoStop}%
\bibitem [{\citenamefont {Spector}(2017)}]{Spector:2016vwo}%
  \BibitemOpen
  \bibfield  {author} {\bibinfo {author} {\bibfnamefont {A.}~\bibnamefont
  {Spector}} (\bibinfo {collaboration} {ALPS}),\ }in\ \href {\doibase
  10.3204/DESY-PROC-2009-03/Spector_Aaron} {\emph {\bibinfo {booktitle} {{12th
  Patras Workshop on Axions, WIMPs and WISPs}}}}\ (\bibinfo {year} {2017})\
  pp.\ \bibinfo {pages} {133--136},\ \Eprint {http://arxiv.org/abs/1611.05863}
  {arXiv:1611.05863 [physics.ins-det]} \BibitemShut {NoStop}%
\bibitem [{\citenamefont {Abe}\ \emph {et~al.}(2018)\citenamefont {Abe} \emph
  {et~al.}}]{Hyper-Kamiokande:2018ofw}%
  \BibitemOpen
  \bibfield  {author} {\bibinfo {author} {\bibfnamefont {K.}~\bibnamefont
  {Abe}} \emph {et~al.} (\bibinfo {collaboration} {Hyper-Kamiokande}),\
  }\href@noop {} {\  (\bibinfo {year} {2018})},\ \Eprint
  {http://arxiv.org/abs/1805.04163} {arXiv:1805.04163 [physics.ins-det]}
  \BibitemShut {NoStop}%
\bibitem [{\citenamefont {Abi}\ \emph {et~al.}(2020{\natexlab{a}})\citenamefont
  {Abi} \emph {et~al.}}]{DUNE:2020ypp}%
  \BibitemOpen
  \bibfield  {author} {\bibinfo {author} {\bibfnamefont {B.}~\bibnamefont
  {Abi}} \emph {et~al.} (\bibinfo {collaboration} {DUNE}),\ }\href@noop {} {\
  (\bibinfo {year} {2020}{\natexlab{a}})},\ \Eprint
  {http://arxiv.org/abs/2002.03005} {arXiv:2002.03005 [hep-ex]} \BibitemShut
  {NoStop}%
\bibitem [{\citenamefont {Abusleme}\ \emph {et~al.}(2021)\citenamefont
  {Abusleme} \emph {et~al.}}]{JUNO:2021vlw}%
  \BibitemOpen
  \bibfield  {author} {\bibinfo {author} {\bibfnamefont {A.}~\bibnamefont
  {Abusleme}} \emph {et~al.} (\bibinfo {collaboration} {JUNO}),\ }\href@noop {}
  {\  (\bibinfo {year} {2021})},\ \Eprint {http://arxiv.org/abs/2104.02565}
  {arXiv:2104.02565 [hep-ex]} \BibitemShut {NoStop}%
\bibitem [{\citenamefont {Cui}\ \emph {et~al.}(2018)\citenamefont {Cui},
  \citenamefont {Pospelov},\ and\ \citenamefont {Pradler}}]{Cui:2017ytb}%
  \BibitemOpen
  \bibfield  {author} {\bibinfo {author} {\bibfnamefont {Y.}~\bibnamefont
  {Cui}}, \bibinfo {author} {\bibfnamefont {M.}~\bibnamefont {Pospelov}}, \
  and\ \bibinfo {author} {\bibfnamefont {J.}~\bibnamefont {Pradler}},\ }\href
  {\doibase 10.1103/PhysRevD.97.103004} {\bibfield  {journal} {\bibinfo
  {journal} {Phys. Rev. D}\ }\textbf {\bibinfo {volume} {97}},\ \bibinfo
  {pages} {103004} (\bibinfo {year} {2018})},\ \Eprint
  {http://arxiv.org/abs/1711.04531} {arXiv:1711.04531 [hep-ph]} \BibitemShut
  {NoStop}%
\bibitem [{\citenamefont {Nikolic}\ \emph {et~al.}(2020)\citenamefont
  {Nikolic}, \citenamefont {Kulkarni},\ and\ \citenamefont
  {Pradler}}]{Nikolic:2020fom}%
  \BibitemOpen
  \bibfield  {author} {\bibinfo {author} {\bibfnamefont {M.}~\bibnamefont
  {Nikolic}}, \bibinfo {author} {\bibfnamefont {S.}~\bibnamefont {Kulkarni}}, \
  and\ \bibinfo {author} {\bibfnamefont {J.}~\bibnamefont {Pradler}},\
  }\href@noop {} {\  (\bibinfo {year} {2020})},\ \Eprint
  {http://arxiv.org/abs/2008.13557} {arXiv:2008.13557 [hep-ph]} \BibitemShut
  {NoStop}%
\bibitem [{\citenamefont {Bondarenko}\ \emph {et~al.}(2021)\citenamefont
  {Bondarenko}, \citenamefont {Boyarsky}, \citenamefont {Nikolic},
  \citenamefont {Pradler},\ and\ \citenamefont
  {Sokolenko}}]{Bondarenko:2020vta}%
  \BibitemOpen
  \bibfield  {author} {\bibinfo {author} {\bibfnamefont {K.}~\bibnamefont
  {Bondarenko}}, \bibinfo {author} {\bibfnamefont {A.}~\bibnamefont
  {Boyarsky}}, \bibinfo {author} {\bibfnamefont {M.}~\bibnamefont {Nikolic}},
  \bibinfo {author} {\bibfnamefont {J.}~\bibnamefont {Pradler}}, \ and\
  \bibinfo {author} {\bibfnamefont {A.}~\bibnamefont {Sokolenko}},\ }\href
  {\doibase 10.1088/1475-7516/2021/03/089} {\bibfield  {journal} {\bibinfo
  {journal} {JCAP}\ }\textbf {\bibinfo {volume} {03}},\ \bibinfo {pages} {089}
  (\bibinfo {year} {2021})},\ \Eprint {http://arxiv.org/abs/2012.09704}
  {arXiv:2012.09704 [astro-ph.CO]} \BibitemShut {NoStop}%
\bibitem [{\citenamefont {Kuo}\ \emph {et~al.}(2021)\citenamefont {Kuo},
  \citenamefont {Pospelov},\ and\ \citenamefont {Pradler}}]{Kuo:2021mtp}%
  \BibitemOpen
  \bibfield  {author} {\bibinfo {author} {\bibfnamefont {J.-L.}\ \bibnamefont
  {Kuo}}, \bibinfo {author} {\bibfnamefont {M.}~\bibnamefont {Pospelov}}, \
  and\ \bibinfo {author} {\bibfnamefont {J.}~\bibnamefont {Pradler}},\ }\href
  {\doibase 10.1103/PhysRevD.103.115030} {\bibfield  {journal} {\bibinfo
  {journal} {Phys. Rev. D}\ }\textbf {\bibinfo {volume} {103}},\ \bibinfo
  {pages} {115030} (\bibinfo {year} {2021})},\ \Eprint
  {http://arxiv.org/abs/2102.08409} {arXiv:2102.08409 [hep-ph]} \BibitemShut
  {NoStop}%
\bibitem [{\citenamefont {Gu}\ \emph {et~al.}(2022)\citenamefont {Gu},
  \citenamefont {Wu},\ and\ \citenamefont {Zhu}}]{Gu:2021lni}%
  \BibitemOpen
  \bibfield  {author} {\bibinfo {author} {\bibfnamefont {Y.}~\bibnamefont
  {Gu}}, \bibinfo {author} {\bibfnamefont {L.}~\bibnamefont {Wu}}, \ and\
  \bibinfo {author} {\bibfnamefont {B.}~\bibnamefont {Zhu}},\ }\href {\doibase
  10.1103/PhysRevD.105.095008} {\bibfield  {journal} {\bibinfo  {journal}
  {Phys. Rev. D}\ }\textbf {\bibinfo {volume} {105}},\ \bibinfo {pages}
  {095008} (\bibinfo {year} {2022})},\ \Eprint
  {http://arxiv.org/abs/2105.07232} {arXiv:2105.07232 [hep-ph]} \BibitemShut
  {NoStop}%
\bibitem [{\citenamefont {Agashe}\ \emph {et~al.}(2014)\citenamefont {Agashe},
  \citenamefont {Cui}, \citenamefont {Necib},\ and\ \citenamefont
  {Thaler}}]{Agashe:2014yua}%
  \BibitemOpen
  \bibfield  {author} {\bibinfo {author} {\bibfnamefont {K.}~\bibnamefont
  {Agashe}}, \bibinfo {author} {\bibfnamefont {Y.}~\bibnamefont {Cui}},
  \bibinfo {author} {\bibfnamefont {L.}~\bibnamefont {Necib}}, \ and\ \bibinfo
  {author} {\bibfnamefont {J.}~\bibnamefont {Thaler}},\ }\href {\doibase
  10.1088/1475-7516/2014/10/062} {\bibfield  {journal} {\bibinfo  {journal}
  {JCAP}\ }\textbf {\bibinfo {volume} {10}},\ \bibinfo {pages} {062} (\bibinfo
  {year} {2014})},\ \Eprint {http://arxiv.org/abs/1405.7370} {arXiv:1405.7370
  [hep-ph]} \BibitemShut {NoStop}%
\bibitem [{\citenamefont {Berger}\ \emph {et~al.}(2015)\citenamefont {Berger},
  \citenamefont {Cui},\ and\ \citenamefont {Zhao}}]{Berger:2014sqa}%
  \BibitemOpen
  \bibfield  {author} {\bibinfo {author} {\bibfnamefont {J.}~\bibnamefont
  {Berger}}, \bibinfo {author} {\bibfnamefont {Y.}~\bibnamefont {Cui}}, \ and\
  \bibinfo {author} {\bibfnamefont {Y.}~\bibnamefont {Zhao}},\ }\href {\doibase
  10.1088/1475-7516/2015/02/005} {\bibfield  {journal} {\bibinfo  {journal}
  {JCAP}\ }\textbf {\bibinfo {volume} {02}},\ \bibinfo {pages} {005} (\bibinfo
  {year} {2015})},\ \Eprint {http://arxiv.org/abs/1410.2246} {arXiv:1410.2246
  [hep-ph]} \BibitemShut {NoStop}%
\bibitem [{\citenamefont {Giudice}\ \emph {et~al.}(2018)\citenamefont
  {Giudice}, \citenamefont {Kim}, \citenamefont {Park},\ and\ \citenamefont
  {Shin}}]{Giudice:2017zke}%
  \BibitemOpen
  \bibfield  {author} {\bibinfo {author} {\bibfnamefont {G.~F.}\ \bibnamefont
  {Giudice}}, \bibinfo {author} {\bibfnamefont {D.}~\bibnamefont {Kim}},
  \bibinfo {author} {\bibfnamefont {J.-C.}\ \bibnamefont {Park}}, \ and\
  \bibinfo {author} {\bibfnamefont {S.}~\bibnamefont {Shin}},\ }\href {\doibase
  10.1016/j.physletb.2018.03.043} {\bibfield  {journal} {\bibinfo  {journal}
  {Phys. Lett. B}\ }\textbf {\bibinfo {volume} {780}},\ \bibinfo {pages} {543}
  (\bibinfo {year} {2018})},\ \Eprint {http://arxiv.org/abs/1712.07126}
  {arXiv:1712.07126 [hep-ph]} \BibitemShut {NoStop}%
\bibitem [{\citenamefont {Berger}\ \emph {et~al.}(2021)\citenamefont {Berger},
  \citenamefont {Cui}, \citenamefont {Graham}, \citenamefont {Necib},
  \citenamefont {Petrillo}, \citenamefont {Stocks}, \citenamefont {Tsai},\ and\
  \citenamefont {Zhao}}]{Berger:2019ttc}%
  \BibitemOpen
  \bibfield  {author} {\bibinfo {author} {\bibfnamefont {J.}~\bibnamefont
  {Berger}}, \bibinfo {author} {\bibfnamefont {Y.}~\bibnamefont {Cui}},
  \bibinfo {author} {\bibfnamefont {M.}~\bibnamefont {Graham}}, \bibinfo
  {author} {\bibfnamefont {L.}~\bibnamefont {Necib}}, \bibinfo {author}
  {\bibfnamefont {G.}~\bibnamefont {Petrillo}}, \bibinfo {author}
  {\bibfnamefont {D.}~\bibnamefont {Stocks}}, \bibinfo {author} {\bibfnamefont
  {Y.-T.}\ \bibnamefont {Tsai}}, \ and\ \bibinfo {author} {\bibfnamefont
  {Y.}~\bibnamefont {Zhao}},\ }\href {\doibase 10.1103/PhysRevD.103.095012}
  {\bibfield  {journal} {\bibinfo  {journal} {Phys. Rev. D}\ }\textbf {\bibinfo
  {volume} {103}},\ \bibinfo {pages} {095012} (\bibinfo {year} {2021})},\
  \Eprint {http://arxiv.org/abs/1912.05558} {arXiv:1912.05558 [hep-ph]}
  \BibitemShut {NoStop}%
\bibitem [{\citenamefont {Huang}\ and\ \citenamefont
  {Zhao}(2014)}]{Huang:2013xfa}%
  \BibitemOpen
  \bibfield  {author} {\bibinfo {author} {\bibfnamefont {J.}~\bibnamefont
  {Huang}}\ and\ \bibinfo {author} {\bibfnamefont {Y.}~\bibnamefont {Zhao}},\
  }\href {\doibase 10.1007/JHEP02(2014)077} {\bibfield  {journal} {\bibinfo
  {journal} {JHEP}\ }\textbf {\bibinfo {volume} {02}},\ \bibinfo {pages} {077}
  (\bibinfo {year} {2014})},\ \Eprint {http://arxiv.org/abs/1312.0011}
  {arXiv:1312.0011 [hep-ph]} \BibitemShut {NoStop}%
\bibitem [{\citenamefont {Kopp}\ \emph {et~al.}(2015)\citenamefont {Kopp},
  \citenamefont {Liu},\ and\ \citenamefont {Wang}}]{Kopp:2015bfa}%
  \BibitemOpen
  \bibfield  {author} {\bibinfo {author} {\bibfnamefont {J.}~\bibnamefont
  {Kopp}}, \bibinfo {author} {\bibfnamefont {J.}~\bibnamefont {Liu}}, \ and\
  \bibinfo {author} {\bibfnamefont {X.-P.}\ \bibnamefont {Wang}},\ }\href
  {\doibase 10.1007/JHEP04(2015)105} {\bibfield  {journal} {\bibinfo  {journal}
  {JHEP}\ }\textbf {\bibinfo {volume} {04}},\ \bibinfo {pages} {105} (\bibinfo
  {year} {2015})},\ \Eprint {http://arxiv.org/abs/1503.02669} {arXiv:1503.02669
  [hep-ph]} \BibitemShut {NoStop}%
\bibitem [{\citenamefont {Kachulis}\ \emph {et~al.}(2018)\citenamefont
  {Kachulis} \emph {et~al.}}]{Super-Kamiokande:2017dch}%
  \BibitemOpen
  \bibfield  {author} {\bibinfo {author} {\bibfnamefont {C.}~\bibnamefont
  {Kachulis}} \emph {et~al.} (\bibinfo {collaboration} {Super-Kamiokande}),\
  }\href {\doibase 10.1103/PhysRevLett.120.221301} {\bibfield  {journal}
  {\bibinfo  {journal} {Phys. Rev. Lett.}\ }\textbf {\bibinfo {volume} {120}},\
  \bibinfo {pages} {221301} (\bibinfo {year} {2018})},\ \Eprint
  {http://arxiv.org/abs/1711.05278} {arXiv:1711.05278 [hep-ex]} \BibitemShut
  {NoStop}%
\bibitem [{\citenamefont {Ha}\ \emph {et~al.}(2019)\citenamefont {Ha} \emph
  {et~al.}}]{COSINE-100:2018ged}%
  \BibitemOpen
  \bibfield  {author} {\bibinfo {author} {\bibfnamefont {C.}~\bibnamefont {Ha}}
  \emph {et~al.} (\bibinfo {collaboration} {COSINE-100}),\ }\href {\doibase
  10.1103/PhysRevLett.122.131802} {\bibfield  {journal} {\bibinfo  {journal}
  {Phys. Rev. Lett.}\ }\textbf {\bibinfo {volume} {122}},\ \bibinfo {pages}
  {131802} (\bibinfo {year} {2019})},\ \Eprint
  {http://arxiv.org/abs/1811.09344} {arXiv:1811.09344 [astro-ph.IM]}
  \BibitemShut {NoStop}%
\bibitem [{\citenamefont {Chatterjee}\ \emph {et~al.}(2018)\citenamefont
  {Chatterjee}, \citenamefont {De~Roeck}, \citenamefont {Kim}, \citenamefont
  {Moghaddam}, \citenamefont {Park}, \citenamefont {Shin}, \citenamefont
  {Whitehead},\ and\ \citenamefont {Yu}}]{Chatterjee:2018mej}%
  \BibitemOpen
  \bibfield  {author} {\bibinfo {author} {\bibfnamefont {A.}~\bibnamefont
  {Chatterjee}}, \bibinfo {author} {\bibfnamefont {A.}~\bibnamefont
  {De~Roeck}}, \bibinfo {author} {\bibfnamefont {D.}~\bibnamefont {Kim}},
  \bibinfo {author} {\bibfnamefont {Z.~G.}\ \bibnamefont {Moghaddam}}, \bibinfo
  {author} {\bibfnamefont {J.-C.}\ \bibnamefont {Park}}, \bibinfo {author}
  {\bibfnamefont {S.}~\bibnamefont {Shin}}, \bibinfo {author} {\bibfnamefont
  {L.~H.}\ \bibnamefont {Whitehead}}, \ and\ \bibinfo {author} {\bibfnamefont
  {J.}~\bibnamefont {Yu}},\ }\href {\doibase 10.1103/PhysRevD.98.075027}
  {\bibfield  {journal} {\bibinfo  {journal} {Phys. Rev. D}\ }\textbf {\bibinfo
  {volume} {98}},\ \bibinfo {pages} {075027} (\bibinfo {year} {2018})},\
  \Eprint {http://arxiv.org/abs/1803.03264} {arXiv:1803.03264 [hep-ph]}
  \BibitemShut {NoStop}%
\bibitem [{\citenamefont {Dent}\ \emph {et~al.}(2020)\citenamefont {Dent},
  \citenamefont {Dutta}, \citenamefont {Newstead},\ and\ \citenamefont
  {Shoemaker}}]{Dent:2019krz}%
  \BibitemOpen
  \bibfield  {author} {\bibinfo {author} {\bibfnamefont {J.~B.}\ \bibnamefont
  {Dent}}, \bibinfo {author} {\bibfnamefont {B.}~\bibnamefont {Dutta}},
  \bibinfo {author} {\bibfnamefont {J.~L.}\ \bibnamefont {Newstead}}, \ and\
  \bibinfo {author} {\bibfnamefont {I.~M.}\ \bibnamefont {Shoemaker}},\ }\href
  {\doibase 10.1103/PhysRevD.101.116007} {\bibfield  {journal} {\bibinfo
  {journal} {Phys. Rev. D}\ }\textbf {\bibinfo {volume} {101}},\ \bibinfo
  {pages} {116007} (\bibinfo {year} {2020})},\ \Eprint
  {http://arxiv.org/abs/1907.03782} {arXiv:1907.03782 [hep-ph]} \BibitemShut
  {NoStop}%
\bibitem [{\citenamefont {Cui}\ \emph {et~al.}(2022)\citenamefont {Cui} \emph
  {et~al.}}]{PandaX-II:2021kai}%
  \BibitemOpen
  \bibfield  {author} {\bibinfo {author} {\bibfnamefont {X.}~\bibnamefont
  {Cui}} \emph {et~al.} (\bibinfo {collaboration} {PandaX-II}),\ }\href
  {\doibase 10.1103/PhysRevLett.128.171801} {\bibfield  {journal} {\bibinfo
  {journal} {Phys. Rev. Lett.}\ }\textbf {\bibinfo {volume} {128}},\ \bibinfo
  {pages} {171801} (\bibinfo {year} {2022})},\ \Eprint
  {http://arxiv.org/abs/2112.08957} {arXiv:2112.08957 [hep-ex]} \BibitemShut
  {NoStop}%
\bibitem [{\citenamefont {Cadamuro}(2012)}]{Cadamuro:2012rm}%
  \BibitemOpen
  \bibfield  {author} {\bibinfo {author} {\bibfnamefont {D.}~\bibnamefont
  {Cadamuro}},\ }\emph {\bibinfo {title} {{Cosmological limits on axions and
  axion-like particles}}},\ \href@noop {} {Ph.D. thesis},\ \bibinfo  {school}
  {Munich U.} (\bibinfo {year} {2012}),\ \Eprint
  {http://arxiv.org/abs/1210.3196} {arXiv:1210.3196 [hep-ph]} \BibitemShut
  {NoStop}%
\bibitem [{\citenamefont {Chang}\ and\ \citenamefont
  {Cui}(2022)}]{Chang:2021afa}%
  \BibitemOpen
  \bibfield  {author} {\bibinfo {author} {\bibfnamefont {C.-F.}\ \bibnamefont
  {Chang}}\ and\ \bibinfo {author} {\bibfnamefont {Y.}~\bibnamefont {Cui}},\
  }\href {\doibase 10.1007/JHEP03(2022)114} {\bibfield  {journal} {\bibinfo
  {journal} {JHEP}\ }\textbf {\bibinfo {volume} {03}},\ \bibinfo {pages} {114}
  (\bibinfo {year} {2022})},\ \Eprint {http://arxiv.org/abs/2106.09746}
  {arXiv:2106.09746 [hep-ph]} \BibitemShut {NoStop}%
\bibitem [{\citenamefont {Hasenkamp}\ and\ \citenamefont
  {Kersten}(2013)}]{Hasenkamp:2012ii}%
  \BibitemOpen
  \bibfield  {author} {\bibinfo {author} {\bibfnamefont {J.}~\bibnamefont
  {Hasenkamp}}\ and\ \bibinfo {author} {\bibfnamefont {J.}~\bibnamefont
  {Kersten}},\ }\href {\doibase 10.1088/1475-7516/2013/08/024} {\bibfield
  {journal} {\bibinfo  {journal} {JCAP}\ }\textbf {\bibinfo {volume} {08}},\
  \bibinfo {pages} {024} (\bibinfo {year} {2013})},\ \Eprint
  {http://arxiv.org/abs/1212.4160} {arXiv:1212.4160 [hep-ph]} \BibitemShut
  {NoStop}%
\bibitem [{\citenamefont {Cui}\ and\ \citenamefont {Huo}(2019)}]{Cui:2018imi}%
  \BibitemOpen
  \bibfield  {author} {\bibinfo {author} {\bibfnamefont {Y.}~\bibnamefont
  {Cui}}\ and\ \bibinfo {author} {\bibfnamefont {R.}~\bibnamefont {Huo}},\
  }\href {\doibase 10.1103/PhysRevD.100.023004} {\bibfield  {journal} {\bibinfo
   {journal} {Phys. Rev. D}\ }\textbf {\bibinfo {volume} {100}},\ \bibinfo
  {pages} {023004} (\bibinfo {year} {2019})},\ \Eprint
  {http://arxiv.org/abs/1805.06451} {arXiv:1805.06451 [astro-ph.CO]}
  \BibitemShut {NoStop}%
\bibitem [{\citenamefont {Navarro}\ \emph {et~al.}(1996)\citenamefont
  {Navarro}, \citenamefont {Frenk},\ and\ \citenamefont
  {White}}]{Navarro:1995iw}%
  \BibitemOpen
  \bibfield  {author} {\bibinfo {author} {\bibfnamefont {J.~F.}\ \bibnamefont
  {Navarro}}, \bibinfo {author} {\bibfnamefont {C.~S.}\ \bibnamefont {Frenk}},
  \ and\ \bibinfo {author} {\bibfnamefont {S.~D.~M.}\ \bibnamefont {White}},\
  }\href {\doibase 10.1086/177173} {\bibfield  {journal} {\bibinfo  {journal}
  {Astrophys. J.}\ }\textbf {\bibinfo {volume} {462}},\ \bibinfo {pages} {563}
  (\bibinfo {year} {1996})},\ \Eprint {http://arxiv.org/abs/astro-ph/9508025}
  {arXiv:astro-ph/9508025} \BibitemShut {NoStop}%
\bibitem [{\citenamefont {Aghanim}\ \emph {et~al.}(2020)\citenamefont {Aghanim}
  \emph {et~al.}}]{Planck:2018vyg}%
  \BibitemOpen
  \bibfield  {author} {\bibinfo {author} {\bibfnamefont {N.}~\bibnamefont
  {Aghanim}} \emph {et~al.} (\bibinfo {collaboration} {Planck}),\ }\href
  {\doibase 10.1051/0004-6361/201833910} {\bibfield  {journal} {\bibinfo
  {journal} {Astron. Astrophys.}\ }\textbf {\bibinfo {volume} {641}},\ \bibinfo
  {pages} {A6} (\bibinfo {year} {2020})},\ \bibinfo {note} {[Erratum:
  Astron.Astrophys. 652, C4 (2021)]},\ \Eprint
  {http://arxiv.org/abs/1807.06209} {arXiv:1807.06209 [astro-ph.CO]}
  \BibitemShut {NoStop}%
\bibitem [{\citenamefont {Chen}\ \emph {et~al.}(2021)\citenamefont {Chen} \emph
  {et~al.}}]{DES:2020mpv}%
  \BibitemOpen
  \bibfield  {author} {\bibinfo {author} {\bibfnamefont {A.}~\bibnamefont
  {Chen}} \emph {et~al.} (\bibinfo {collaboration} {DES}),\ }\href {\doibase
  10.1103/PhysRevD.103.123528} {\bibfield  {journal} {\bibinfo  {journal}
  {Phys. Rev. D}\ }\textbf {\bibinfo {volume} {103}},\ \bibinfo {pages}
  {123528} (\bibinfo {year} {2021})},\ \Eprint
  {http://arxiv.org/abs/2011.04606} {arXiv:2011.04606 [astro-ph.CO]}
  \BibitemShut {NoStop}%
\bibitem [{\citenamefont {Enqvist}\ \emph {et~al.}(2015)\citenamefont
  {Enqvist}, \citenamefont {Nadathur}, \citenamefont {Sekiguchi},\ and\
  \citenamefont {Takahashi}}]{Enqvist:2015ara}%
  \BibitemOpen
  \bibfield  {author} {\bibinfo {author} {\bibfnamefont {K.}~\bibnamefont
  {Enqvist}}, \bibinfo {author} {\bibfnamefont {S.}~\bibnamefont {Nadathur}},
  \bibinfo {author} {\bibfnamefont {T.}~\bibnamefont {Sekiguchi}}, \ and\
  \bibinfo {author} {\bibfnamefont {T.}~\bibnamefont {Takahashi}},\ }\href
  {\doibase 10.1088/1475-7516/2015/09/067} {\bibfield  {journal} {\bibinfo
  {journal} {JCAP}\ }\textbf {\bibinfo {volume} {09}},\ \bibinfo {pages} {067}
  (\bibinfo {year} {2015})},\ \Eprint {http://arxiv.org/abs/1505.05511}
  {arXiv:1505.05511 [astro-ph.CO]} \BibitemShut {NoStop}%
\bibitem [{\citenamefont {Poulin}\ \emph {et~al.}(2016)\citenamefont {Poulin},
  \citenamefont {Serpico},\ and\ \citenamefont {Lesgourgues}}]{Poulin:2016nat}%
  \BibitemOpen
  \bibfield  {author} {\bibinfo {author} {\bibfnamefont {V.}~\bibnamefont
  {Poulin}}, \bibinfo {author} {\bibfnamefont {P.~D.}\ \bibnamefont {Serpico}},
  \ and\ \bibinfo {author} {\bibfnamefont {J.}~\bibnamefont {Lesgourgues}},\
  }\href {\doibase 10.1088/1475-7516/2016/08/036} {\bibfield  {journal}
  {\bibinfo  {journal} {JCAP}\ }\textbf {\bibinfo {volume} {08}},\ \bibinfo
  {pages} {036} (\bibinfo {year} {2016})},\ \Eprint
  {http://arxiv.org/abs/1606.02073} {arXiv:1606.02073 [astro-ph.CO]}
  \BibitemShut {NoStop}%
\bibitem [{\citenamefont {Nygaard}\ \emph {et~al.}(2021)\citenamefont
  {Nygaard}, \citenamefont {Tram},\ and\ \citenamefont
  {Hannestad}}]{Nygaard:2020sow}%
  \BibitemOpen
  \bibfield  {author} {\bibinfo {author} {\bibfnamefont {A.}~\bibnamefont
  {Nygaard}}, \bibinfo {author} {\bibfnamefont {T.}~\bibnamefont {Tram}}, \
  and\ \bibinfo {author} {\bibfnamefont {S.}~\bibnamefont {Hannestad}},\ }\href
  {\doibase 10.1088/1475-7516/2021/05/017} {\bibfield  {journal} {\bibinfo
  {journal} {JCAP}\ }\textbf {\bibinfo {volume} {05}},\ \bibinfo {pages} {017}
  (\bibinfo {year} {2021})},\ \Eprint {http://arxiv.org/abs/2011.01632}
  {arXiv:2011.01632 [astro-ph.CO]} \BibitemShut {NoStop}%
\bibitem [{\citenamefont {Simon}\ \emph {et~al.}(2022)\citenamefont {Simon},
  \citenamefont {Franco~Abell\'an}, \citenamefont {Du}, \citenamefont
  {Poulin},\ and\ \citenamefont {Tsai}}]{Simon:2022ftd}%
  \BibitemOpen
  \bibfield  {author} {\bibinfo {author} {\bibfnamefont {T.}~\bibnamefont
  {Simon}}, \bibinfo {author} {\bibfnamefont {G.}~\bibnamefont
  {Franco~Abell\'an}}, \bibinfo {author} {\bibfnamefont {P.}~\bibnamefont
  {Du}}, \bibinfo {author} {\bibfnamefont {V.}~\bibnamefont {Poulin}}, \ and\
  \bibinfo {author} {\bibfnamefont {Y.}~\bibnamefont {Tsai}},\ }\href {\doibase
  10.1103/PhysRevD.106.023516} {\bibfield  {journal} {\bibinfo  {journal}
  {Phys. Rev. D}\ }\textbf {\bibinfo {volume} {106}},\ \bibinfo {pages}
  {023516} (\bibinfo {year} {2022})},\ \Eprint
  {http://arxiv.org/abs/2203.07440} {arXiv:2203.07440 [astro-ph.CO]}
  \BibitemShut {NoStop}%
\bibitem [{\citenamefont {Cadamuro}\ and\ \citenamefont
  {Redondo}(2012)}]{Cadamuro:2011fd}%
  \BibitemOpen
  \bibfield  {author} {\bibinfo {author} {\bibfnamefont {D.}~\bibnamefont
  {Cadamuro}}\ and\ \bibinfo {author} {\bibfnamefont {J.}~\bibnamefont
  {Redondo}},\ }\href {\doibase 10.1088/1475-7516/2012/02/032} {\bibfield
  {journal} {\bibinfo  {journal} {JCAP}\ }\textbf {\bibinfo {volume} {02}},\
  \bibinfo {pages} {032} (\bibinfo {year} {2012})},\ \Eprint
  {http://arxiv.org/abs/1110.2895} {arXiv:1110.2895 [hep-ph]} \BibitemShut
  {NoStop}%
\bibitem [{\citenamefont {Cohen}\ \emph {et~al.}(2017)\citenamefont {Cohen},
  \citenamefont {Murase}, \citenamefont {Rodd}, \citenamefont {Safdi},\ and\
  \citenamefont {Soreq}}]{Cohen:2016uyg}%
  \BibitemOpen
  \bibfield  {author} {\bibinfo {author} {\bibfnamefont {T.}~\bibnamefont
  {Cohen}}, \bibinfo {author} {\bibfnamefont {K.}~\bibnamefont {Murase}},
  \bibinfo {author} {\bibfnamefont {N.~L.}\ \bibnamefont {Rodd}}, \bibinfo
  {author} {\bibfnamefont {B.~R.}\ \bibnamefont {Safdi}}, \ and\ \bibinfo
  {author} {\bibfnamefont {Y.}~\bibnamefont {Soreq}},\ }\href {\doibase
  10.1103/PhysRevLett.119.021102} {\bibfield  {journal} {\bibinfo  {journal}
  {Phys. Rev. Lett.}\ }\textbf {\bibinfo {volume} {119}},\ \bibinfo {pages}
  {021102} (\bibinfo {year} {2017})},\ \Eprint
  {http://arxiv.org/abs/1612.05638} {arXiv:1612.05638 [hep-ph]} \BibitemShut
  {NoStop}%
\bibitem [{\citenamefont {Foster}\ \emph {et~al.}(2021)\citenamefont {Foster},
  \citenamefont {Kongsore}, \citenamefont {Dessert}, \citenamefont {Park},
  \citenamefont {Rodd}, \citenamefont {Cranmer},\ and\ \citenamefont
  {Safdi}}]{Foster:2021ngm}%
  \BibitemOpen
  \bibfield  {author} {\bibinfo {author} {\bibfnamefont {J.~W.}\ \bibnamefont
  {Foster}}, \bibinfo {author} {\bibfnamefont {M.}~\bibnamefont {Kongsore}},
  \bibinfo {author} {\bibfnamefont {C.}~\bibnamefont {Dessert}}, \bibinfo
  {author} {\bibfnamefont {Y.}~\bibnamefont {Park}}, \bibinfo {author}
  {\bibfnamefont {N.~L.}\ \bibnamefont {Rodd}}, \bibinfo {author}
  {\bibfnamefont {K.}~\bibnamefont {Cranmer}}, \ and\ \bibinfo {author}
  {\bibfnamefont {B.~R.}\ \bibnamefont {Safdi}},\ }\href {\doibase
  10.1103/PhysRevLett.127.051101} {\bibfield  {journal} {\bibinfo  {journal}
  {Phys. Rev. Lett.}\ }\textbf {\bibinfo {volume} {127}},\ \bibinfo {pages}
  {051101} (\bibinfo {year} {2021})},\ \Eprint
  {http://arxiv.org/abs/2102.02207} {arXiv:2102.02207 [astro-ph.CO]}
  \BibitemShut {NoStop}%
\bibitem [{\citenamefont {Roach}\ \emph {et~al.}(2022)\citenamefont {Roach},
  \citenamefont {Rossland}, \citenamefont {Ng}, \citenamefont {Perez},
  \citenamefont {Beacom}, \citenamefont {Grefenstette}, \citenamefont
  {Horiuchi}, \citenamefont {Krivonos},\ and\ \citenamefont
  {Wik}}]{Roach:2022lgo}%
  \BibitemOpen
  \bibfield  {author} {\bibinfo {author} {\bibfnamefont {B.~M.}\ \bibnamefont
  {Roach}}, \bibinfo {author} {\bibfnamefont {S.}~\bibnamefont {Rossland}},
  \bibinfo {author} {\bibfnamefont {K.~C.~Y.}\ \bibnamefont {Ng}}, \bibinfo
  {author} {\bibfnamefont {K.}~\bibnamefont {Perez}}, \bibinfo {author}
  {\bibfnamefont {J.~F.}\ \bibnamefont {Beacom}}, \bibinfo {author}
  {\bibfnamefont {B.~W.}\ \bibnamefont {Grefenstette}}, \bibinfo {author}
  {\bibfnamefont {S.}~\bibnamefont {Horiuchi}}, \bibinfo {author}
  {\bibfnamefont {R.}~\bibnamefont {Krivonos}}, \ and\ \bibinfo {author}
  {\bibfnamefont {D.~R.}\ \bibnamefont {Wik}},\ }\href@noop {} {\  (\bibinfo
  {year} {2022})},\ \Eprint {http://arxiv.org/abs/2207.04572} {arXiv:2207.04572
  [astro-ph.HE]} \BibitemShut {NoStop}%
\bibitem [{\citenamefont {Ayala}\ \emph {et~al.}(2014)\citenamefont {Ayala},
  \citenamefont {Dom\'\i{}nguez}, \citenamefont {Giannotti}, \citenamefont
  {Mirizzi},\ and\ \citenamefont {Straniero}}]{Ayala:2014pea}%
  \BibitemOpen
  \bibfield  {author} {\bibinfo {author} {\bibfnamefont {A.}~\bibnamefont
  {Ayala}}, \bibinfo {author} {\bibfnamefont {I.}~\bibnamefont
  {Dom\'\i{}nguez}}, \bibinfo {author} {\bibfnamefont {M.}~\bibnamefont
  {Giannotti}}, \bibinfo {author} {\bibfnamefont {A.}~\bibnamefont {Mirizzi}},
  \ and\ \bibinfo {author} {\bibfnamefont {O.}~\bibnamefont {Straniero}},\
  }\href {\doibase 10.1103/PhysRevLett.113.191302} {\bibfield  {journal}
  {\bibinfo  {journal} {Phys. Rev. Lett.}\ }\textbf {\bibinfo {volume} {113}},\
  \bibinfo {pages} {191302} (\bibinfo {year} {2014})},\ \Eprint
  {http://arxiv.org/abs/1406.6053} {arXiv:1406.6053 [astro-ph.SR]} \BibitemShut
  {NoStop}%
\bibitem [{\citenamefont {Dolan}\ \emph {et~al.}(2022)\citenamefont {Dolan},
  \citenamefont {Hiskens},\ and\ \citenamefont {Volkas}}]{Dolan:2022kul}%
  \BibitemOpen
  \bibfield  {author} {\bibinfo {author} {\bibfnamefont {M.~J.}\ \bibnamefont
  {Dolan}}, \bibinfo {author} {\bibfnamefont {F.~J.}\ \bibnamefont {Hiskens}},
  \ and\ \bibinfo {author} {\bibfnamefont {R.~R.}\ \bibnamefont {Volkas}},\
  }\href@noop {} {\  (\bibinfo {year} {2022})},\ \Eprint
  {http://arxiv.org/abs/2207.03102} {arXiv:2207.03102 [hep-ph]} \BibitemShut
  {NoStop}%
\bibitem [{\citenamefont {O'Hare}(2021)}]{githublimits}%
  \BibitemOpen
  \bibfield  {author} {\bibinfo {author} {\bibfnamefont {C.}~\bibnamefont
  {O'Hare}},\ }\href@noop {} {\enquote {\bibinfo {title} {Axionlimits},}\
  }\bibinfo {howpublished} {\url{https://github.com/cajohare/AxionLimits}}
  (\bibinfo {year} {2021})\BibitemShut {NoStop}%
\bibitem [{\citenamefont {DeRocco}\ \emph {et~al.}(2020)\citenamefont
  {DeRocco}, \citenamefont {Graham},\ and\ \citenamefont
  {Rajendran}}]{DeRocco:2020xdt}%
  \BibitemOpen
  \bibfield  {author} {\bibinfo {author} {\bibfnamefont {W.}~\bibnamefont
  {DeRocco}}, \bibinfo {author} {\bibfnamefont {P.~W.}\ \bibnamefont {Graham}},
  \ and\ \bibinfo {author} {\bibfnamefont {S.}~\bibnamefont {Rajendran}},\
  }\href {\doibase 10.1103/PhysRevD.102.075015} {\bibfield  {journal} {\bibinfo
   {journal} {Phys. Rev. D}\ }\textbf {\bibinfo {volume} {102}},\ \bibinfo
  {pages} {075015} (\bibinfo {year} {2020})},\ \Eprint
  {http://arxiv.org/abs/2006.15112} {arXiv:2006.15112 [hep-ph]} \BibitemShut
  {NoStop}%
\bibitem [{\citenamefont {Bar}\ \emph {et~al.}(2020)\citenamefont {Bar},
  \citenamefont {Blum},\ and\ \citenamefont {D'Amico}}]{Bar:2019ifz}%
  \BibitemOpen
  \bibfield  {author} {\bibinfo {author} {\bibfnamefont {N.}~\bibnamefont
  {Bar}}, \bibinfo {author} {\bibfnamefont {K.}~\bibnamefont {Blum}}, \ and\
  \bibinfo {author} {\bibfnamefont {G.}~\bibnamefont {D'Amico}},\ }\href
  {\doibase 10.1103/PhysRevD.101.123025} {\bibfield  {journal} {\bibinfo
  {journal} {Phys. Rev. D}\ }\textbf {\bibinfo {volume} {101}},\ \bibinfo
  {pages} {123025} (\bibinfo {year} {2020})},\ \Eprint
  {http://arxiv.org/abs/1907.05020} {arXiv:1907.05020 [hep-ph]} \BibitemShut
  {NoStop}%
\bibitem [{\citenamefont {Bauer}\ \emph {et~al.}(2017)\citenamefont {Bauer},
  \citenamefont {Neubert},\ and\ \citenamefont {Thamm}}]{Bauer:2017ris}%
  \BibitemOpen
  \bibfield  {author} {\bibinfo {author} {\bibfnamefont {M.}~\bibnamefont
  {Bauer}}, \bibinfo {author} {\bibfnamefont {M.}~\bibnamefont {Neubert}}, \
  and\ \bibinfo {author} {\bibfnamefont {A.}~\bibnamefont {Thamm}},\ }\href
  {\doibase 10.1007/JHEP12(2017)044} {\bibfield  {journal} {\bibinfo  {journal}
  {JHEP}\ }\textbf {\bibinfo {volume} {12}},\ \bibinfo {pages} {044} (\bibinfo
  {year} {2017})},\ \Eprint {http://arxiv.org/abs/1708.00443} {arXiv:1708.00443
  [hep-ph]} \BibitemShut {NoStop}%
\bibitem [{\citenamefont {Capozzi}\ and\ \citenamefont
  {Raffelt}(2020)}]{Capozzi:2020cbu}%
  \BibitemOpen
  \bibfield  {author} {\bibinfo {author} {\bibfnamefont {F.}~\bibnamefont
  {Capozzi}}\ and\ \bibinfo {author} {\bibfnamefont {G.}~\bibnamefont
  {Raffelt}},\ }\href {\doibase 10.1103/PhysRevD.102.083007} {\bibfield
  {journal} {\bibinfo  {journal} {Phys. Rev. D}\ }\textbf {\bibinfo {volume}
  {102}},\ \bibinfo {pages} {083007} (\bibinfo {year} {2020})},\ \Eprint
  {http://arxiv.org/abs/2007.03694} {arXiv:2007.03694 [astro-ph.SR]}
  \BibitemShut {NoStop}%
\bibitem [{\citenamefont {Bays}\ \emph {et~al.}(2012)\citenamefont {Bays} \emph
  {et~al.}}]{Super-Kamiokande:2011lwo}%
  \BibitemOpen
  \bibfield  {author} {\bibinfo {author} {\bibfnamefont {K.}~\bibnamefont
  {Bays}} \emph {et~al.} (\bibinfo {collaboration} {Super-Kamiokande}),\ }\href
  {\doibase 10.1103/PhysRevD.85.052007} {\bibfield  {journal} {\bibinfo
  {journal} {Phys. Rev. D}\ }\textbf {\bibinfo {volume} {85}},\ \bibinfo
  {pages} {052007} (\bibinfo {year} {2012})},\ \Eprint
  {http://arxiv.org/abs/1111.5031} {arXiv:1111.5031 [hep-ex]} \BibitemShut
  {NoStop}%
\bibitem [{\citenamefont {Zyla}\ \emph {et~al.}(2020)\citenamefont {Zyla} \emph
  {et~al.}}]{ParticleDataGroup:2020ssz}%
  \BibitemOpen
  \bibfield  {author} {\bibinfo {author} {\bibfnamefont {P.~A.}\ \bibnamefont
  {Zyla}} \emph {et~al.} (\bibinfo {collaboration} {Particle Data Group}),\
  }\href {\doibase 10.1093/ptep/ptaa104} {\bibfield  {journal} {\bibinfo
  {journal} {PTEP}\ }\textbf {\bibinfo {volume} {2020}},\ \bibinfo {pages}
  {083C01} (\bibinfo {year} {2020})}\BibitemShut {NoStop}%
\bibitem [{\citenamefont {Abe}\ \emph {et~al.}(2011)\citenamefont {Abe} \emph
  {et~al.}}]{Abe:2011ts}%
  \BibitemOpen
  \bibfield  {author} {\bibinfo {author} {\bibfnamefont {K.}~\bibnamefont
  {Abe}} \emph {et~al.},\ }\href@noop {} {\  (\bibinfo {year} {2011})},\
  \Eprint {http://arxiv.org/abs/1109.3262} {arXiv:1109.3262 [hep-ex]}
  \BibitemShut {NoStop}%
\bibitem [{\citenamefont {Kearns}\ \emph {et~al.}(2013)\citenamefont {Kearns}
  \emph {et~al.}}]{Hyper-KamiokandeWorkingGroup:2013hcb}%
  \BibitemOpen
  \bibfield  {author} {\bibinfo {author} {\bibfnamefont {E.}~\bibnamefont
  {Kearns}} \emph {et~al.} (\bibinfo {collaboration} {Hyper-Kamiokande Working
  Group}),\ }in\ \href@noop {} {\emph {\bibinfo {booktitle} {{Community Summer
  Study 2013}: {Snowmass on the Mississippi}}}}\ (\bibinfo {year} {2013})\
  \Eprint {http://arxiv.org/abs/1309.0184} {arXiv:1309.0184 [hep-ex]}
  \BibitemShut {NoStop}%
\bibitem [{\citenamefont {Aartsen}\ \emph {et~al.}(2015)\citenamefont {Aartsen}
  \emph {et~al.}}]{IceCube:2015mgt}%
  \BibitemOpen
  \bibfield  {author} {\bibinfo {author} {\bibfnamefont {M.~G.}\ \bibnamefont
  {Aartsen}} \emph {et~al.} (\bibinfo {collaboration} {IceCube}),\ }\href
  {\doibase 10.1103/PhysRevD.91.122004} {\bibfield  {journal} {\bibinfo
  {journal} {Phys. Rev. D}\ }\textbf {\bibinfo {volume} {91}},\ \bibinfo
  {pages} {122004} (\bibinfo {year} {2015})},\ \Eprint
  {http://arxiv.org/abs/1504.03753} {arXiv:1504.03753 [astro-ph.HE]}
  \BibitemShut {NoStop}%
\bibitem [{\citenamefont {Aartsen}\ \emph {et~al.}(2013)\citenamefont {Aartsen}
  \emph {et~al.}}]{IceCube:2012jwm}%
  \BibitemOpen
  \bibfield  {author} {\bibinfo {author} {\bibfnamefont {M.~G.}\ \bibnamefont
  {Aartsen}} \emph {et~al.} (\bibinfo {collaboration} {IceCube}),\ }\href
  {\doibase 10.1103/PhysRevLett.110.151105} {\bibfield  {journal} {\bibinfo
  {journal} {Phys. Rev. Lett.}\ }\textbf {\bibinfo {volume} {110}},\ \bibinfo
  {pages} {151105} (\bibinfo {year} {2013})},\ \Eprint
  {http://arxiv.org/abs/1212.4760} {arXiv:1212.4760 [hep-ex]} \BibitemShut
  {NoStop}%
\bibitem [{\citenamefont {Honda}\ \emph {et~al.}(2007)\citenamefont {Honda},
  \citenamefont {Kajita}, \citenamefont {Kasahara}, \citenamefont
  {Midorikawa},\ and\ \citenamefont {Sanuki}}]{Honda:2006qj}%
  \BibitemOpen
  \bibfield  {author} {\bibinfo {author} {\bibfnamefont {M.}~\bibnamefont
  {Honda}}, \bibinfo {author} {\bibfnamefont {T.}~\bibnamefont {Kajita}},
  \bibinfo {author} {\bibfnamefont {K.}~\bibnamefont {Kasahara}}, \bibinfo
  {author} {\bibfnamefont {S.}~\bibnamefont {Midorikawa}}, \ and\ \bibinfo
  {author} {\bibfnamefont {T.}~\bibnamefont {Sanuki}},\ }\href {\doibase
  10.1103/PhysRevD.75.043006} {\bibfield  {journal} {\bibinfo  {journal} {Phys.
  Rev. D}\ }\textbf {\bibinfo {volume} {75}},\ \bibinfo {pages} {043006}
  (\bibinfo {year} {2007})},\ \Eprint {http://arxiv.org/abs/astro-ph/0611418}
  {arXiv:astro-ph/0611418} \BibitemShut {NoStop}%
\bibitem [{\citenamefont {Ishihara}(2021)}]{Ishihara:2019aao}%
  \BibitemOpen
  \bibfield  {author} {\bibinfo {author} {\bibfnamefont {A.}~\bibnamefont
  {Ishihara}} (\bibinfo {collaboration} {IceCube}),\ }\href {\doibase
  10.22323/1.358.1031} {\bibfield  {journal} {\bibinfo  {journal} {PoS}\
  }\textbf {\bibinfo {volume} {ICRC2019}},\ \bibinfo {pages} {1031} (\bibinfo
  {year} {2021})},\ \Eprint {http://arxiv.org/abs/1908.09441} {arXiv:1908.09441
  [astro-ph.HE]} \BibitemShut {NoStop}%
\bibitem [{\citenamefont {Abi}\ \emph {et~al.}(2020{\natexlab{b}})\citenamefont
  {Abi} \emph {et~al.}}]{DUNE:2020txw}%
  \BibitemOpen
  \bibfield  {author} {\bibinfo {author} {\bibfnamefont {B.}~\bibnamefont
  {Abi}} \emph {et~al.} (\bibinfo {collaboration} {DUNE}),\ }\href {\doibase
  10.1088/1748-0221/15/08/T08010} {\bibfield  {journal} {\bibinfo  {journal}
  {JINST}\ }\textbf {\bibinfo {volume} {15}},\ \bibinfo {pages} {T08010}
  (\bibinfo {year} {2020}{\natexlab{b}})},\ \Eprint
  {http://arxiv.org/abs/2002.03010} {arXiv:2002.03010 [physics.ins-det]}
  \BibitemShut {NoStop}%
\bibitem [{\citenamefont {Abi}\ \emph {et~al.}(2018)\citenamefont {Abi} \emph
  {et~al.}}]{DUNE:2018mlo}%
  \BibitemOpen
  \bibfield  {author} {\bibinfo {author} {\bibfnamefont {B.}~\bibnamefont
  {Abi}} \emph {et~al.} (\bibinfo {collaboration} {DUNE}),\ }\href@noop {} {\
  (\bibinfo {year} {2018})},\ \Eprint {http://arxiv.org/abs/1807.10340}
  {arXiv:1807.10340 [physics.ins-det]} \BibitemShut {NoStop}%
\bibitem [{\citenamefont {Abusleme}\ \emph {et~al.}(2022)\citenamefont
  {Abusleme} \emph {et~al.}}]{JUNO:2022lpc}%
  \BibitemOpen
  \bibfield  {author} {\bibinfo {author} {\bibfnamefont {A.}~\bibnamefont
  {Abusleme}} \emph {et~al.} (\bibinfo {collaboration} {JUNO}),\ }\href@noop {}
  {\  (\bibinfo {year} {2022})},\ \Eprint {http://arxiv.org/abs/2205.08830}
  {arXiv:2205.08830 [hep-ex]} \BibitemShut {NoStop}%
\bibitem [{\citenamefont {Rebber}\ \emph {et~al.}(2021)\citenamefont {Rebber},
  \citenamefont {Ludhova}, \citenamefont {Wonsak},\ and\ \citenamefont
  {Xu}}]{Rebber:2020xfi}%
  \BibitemOpen
  \bibfield  {author} {\bibinfo {author} {\bibfnamefont {H.}~\bibnamefont
  {Rebber}}, \bibinfo {author} {\bibfnamefont {L.}~\bibnamefont {Ludhova}},
  \bibinfo {author} {\bibfnamefont {B.~S.}\ \bibnamefont {Wonsak}}, \ and\
  \bibinfo {author} {\bibfnamefont {Y.}~\bibnamefont {Xu}},\ }\href {\doibase
  10.1088/1748-0221/16/01/P01016} {\bibfield  {journal} {\bibinfo  {journal}
  {JINST}\ }\textbf {\bibinfo {volume} {16}},\ \bibinfo {pages} {P01016}
  (\bibinfo {year} {2021})},\ \Eprint {http://arxiv.org/abs/2007.02687}
  {arXiv:2007.02687 [physics.ins-det]} \BibitemShut {NoStop}%
\bibitem [{\citenamefont {Albert}\ \emph {et~al.}(2021)\citenamefont {Albert}
  \emph {et~al.}}]{ANTARES:2021cwc}%
  \BibitemOpen
  \bibfield  {author} {\bibinfo {author} {\bibfnamefont {A.}~\bibnamefont
  {Albert}} \emph {et~al.} (\bibinfo {collaboration} {ANTARES}),\ }\href
  {\doibase 10.1016/j.physletb.2021.136228} {\bibfield  {journal} {\bibinfo
  {journal} {Phys. Lett. B}\ }\textbf {\bibinfo {volume} {816}},\ \bibinfo
  {pages} {136228} (\bibinfo {year} {2021})},\ \Eprint
  {http://arxiv.org/abs/2101.12170} {arXiv:2101.12170 [hep-ex]} \BibitemShut
  {NoStop}%
\bibitem [{\citenamefont {Adrian-Martinez}\ \emph {et~al.}(2016)\citenamefont
  {Adrian-Martinez} \emph {et~al.}}]{KM3Net:2016zxf}%
  \BibitemOpen
  \bibfield  {author} {\bibinfo {author} {\bibfnamefont {S.}~\bibnamefont
  {Adrian-Martinez}} \emph {et~al.} (\bibinfo {collaboration} {KM3Net}),\
  }\href {\doibase 10.1088/0954-3899/43/8/084001} {\bibfield  {journal}
  {\bibinfo  {journal} {J. Phys. G}\ }\textbf {\bibinfo {volume} {43}},\
  \bibinfo {pages} {084001} (\bibinfo {year} {2016})},\ \Eprint
  {http://arxiv.org/abs/1601.07459} {arXiv:1601.07459 [astro-ph.IM]}
  \BibitemShut {NoStop}%
\bibitem [{\citenamefont {Shirai}(2017)}]{Shirai:2017jyz}%
  \BibitemOpen
  \bibfield  {author} {\bibinfo {author} {\bibfnamefont {J.}~\bibnamefont
  {Shirai}} (\bibinfo {collaboration} {KamLAND-Zen}),\ }\href {\doibase
  10.1088/1742-6596/888/1/012031} {\bibfield  {journal} {\bibinfo  {journal}
  {J. Phys. Conf. Ser.}\ }\textbf {\bibinfo {volume} {888}},\ \bibinfo {pages}
  {012031} (\bibinfo {year} {2017})}\BibitemShut {NoStop}%
\bibitem [{\citenamefont {Kim}\ and\ \citenamefont {Tsai}(1973)}]{Kim:1973he}%
  \BibitemOpen
  \bibfield  {author} {\bibinfo {author} {\bibfnamefont {K.~J.}\ \bibnamefont
  {Kim}}\ and\ \bibinfo {author} {\bibfnamefont {Y.-S.}\ \bibnamefont {Tsai}},\
  }\href {\doibase 10.1103/PhysRevD.8.3109} {\bibfield  {journal} {\bibinfo
  {journal} {Phys. Rev. D}\ }\textbf {\bibinfo {volume} {8}},\ \bibinfo {pages}
  {3109} (\bibinfo {year} {1973})}\BibitemShut {NoStop}%
\bibitem [{\citenamefont {Tsai}(1974)}]{Tsai:1973py}%
  \BibitemOpen
  \bibfield  {author} {\bibinfo {author} {\bibfnamefont {Y.-S.}\ \bibnamefont
  {Tsai}},\ }\href {\doibase 10.1103/RevModPhys.46.815} {\bibfield  {journal}
  {\bibinfo  {journal} {Rev. Mod. Phys.}\ }\textbf {\bibinfo {volume} {46}},\
  \bibinfo {pages} {815} (\bibinfo {year} {1974})},\ \bibinfo {note} {[Erratum:
  Rev.Mod.Phys. 49, 421--423 (1977)]}\BibitemShut {NoStop}%
\bibitem [{\citenamefont {Helm}(1956)}]{Helm:1956zz}%
  \BibitemOpen
  \bibfield  {author} {\bibinfo {author} {\bibfnamefont {R.~H.}\ \bibnamefont
  {Helm}},\ }\href {\doibase 10.1103/PhysRev.104.1466} {\bibfield  {journal}
  {\bibinfo  {journal} {Phys. Rev.}\ }\textbf {\bibinfo {volume} {104}},\
  \bibinfo {pages} {1466} (\bibinfo {year} {1956})}\BibitemShut {NoStop}%
\end{thebibliography}%

\end{document}